\documentclass[conf]{new-aiaa}
\usepackage[utf8]{inputenc}

\usepackage{graphicx}
\usepackage[version=4]{mhchem}
\usepackage{siunitx}
\usepackage{longtable,tabularx}

\usepackage{nomencl}
\usepackage{ifthen}

\usepackage{diagbox}
\usepackage{booktabs}
\usepackage{subcaption}

\usepackage{latexsym}%
\usepackage{graphicx}%
\usepackage{verbatim}%

\usepackage{tabularx}

\usepackage{color}

\usepackage{amsmath, amssymb, amsfonts, dsfont}

\usepackage{bm}
\usepackage{amsmath,amssymb,mathrsfs,amsfonts,dsfont} 
\usepackage{cleveref}

\newcommand{\bchi}{\ensuremath{\bm{\chi}}}

\DeclareMathOperator*{\argmax}{argmax}

\newcommand{\overbar}[1]{\mkern 1.2mu\overline{\mkern-1.2mu#1\mkern-1.2mu}\mkern
1.2mu}

\newcommand{\revision}[1]{\textcolor{black}{#1}}

\title{\revision{Surrogate Modeling of Aerodynamic Simulations for
Multiple Operating Conditions Using Machine Learning}}

\author{Romain Dupuis,\footnote{PhD. Student, Embedded Systems Department, 118 route de Narbonne, Toulouse.}
Jean-Christophe Jouhaud,\footnote{Senior researcher, CFD Team - AAM Group, 42
Avenue Gaspard Coriolis, Toulouse.}}
\affil{IRT Saint Exup\'{e}ry, Toulouse, 31 432, France \\ CERFACS,
Toulouse, 31 057, France}
\author{Pierre Sagaut\footnote{Professor and Director, Aix
Marseille Univ, CNRS, Centrale Marseille, M2P2 UMR 7340, 13451 Marseille cedex,
France.}} \affil{Aix Marseille Univ, CNRS, Centrale Marseille, M2P2 UMR 7340, 13451
Marseille cedex, France}

\begin{document}

\maketitle

\begin{abstract}
This article presents an original methodology for the prediction of steady
turbulent aerodynamic fields. Due to the important computational cost of
high-fidelity aerodynamic simulations, a surrogate model is employed to cope
with the significant variations of several inflow conditions. Specifically, the
Local Decomposition Method presented in this paper has been derived to capture
nonlinear behaviors resulting from the presence of continuous and discontinuous
signals. A combination of unsupervised and supervised learning algorithms is
coupled with a physical criterion. It decomposes automatically the input
parameter space, from a limited number of high-fidelity simulations, into
subspaces. These latter correspond to different flow regimes. A measure of
entropy identifies the subspace with the expected strongest non-linear behavior
allowing to perform an active resampling on this low-dimensional
structure. Local reduced-order models are built on each subspace using Proper
Orthogonal Decomposition coupled with a multivariate interpolation tool. The methodology is
assessed on the turbulent two-dimensional flow around
the RAE2822 transonic airfoil. It exhibits a significant improvement in term of
prediction accuracy for the Local Decomposition Method compared with the
classical method of surrogate modeling for cases with different flow regimes.
\end{abstract}

\section*{Nomenclature}

{\renewcommand\arraystretch{1.0}
\noindent\begin{longtable*}{@{}l @{\quad=\quad} l@{}}
$\bm{A}$  & matrix of the reduced coordinates  \\
$a_k$  & $k$-th reduced coordinate  \\
$\bm{B}$  & matrix of the reduced coordinates of the sensor  \\
$b_k$  & $k$-th reduced coordinate of the sensor  \\
$C$  & chord length  \\
$C_k$  & $k$-th cluster  \\
$C_f$  & friction coefficient  \\
$C_p$  & pressure coefficient  \\
$d$  & dimension of the quantity of interest  \\
$E$  & averaged normalized error  \\
$\bm{f}$  & high fidelity model  \\
$g$  & acceleration due to the gravity or normal distribution\\
$H$  & global entropy  \\
$h$  & altitude  \\
$L$  & temperature lapse rate \\
$\bm{l}$  & latent function matrix  \\
$l$  & latent function  \\
$M$  & Mach number  \\
$m$  & number of predictions  \\
$\mathcal{N}$ & Gaussian probability distribution \\
$n$  & number of training samples  \\

$p$  & dimension of an input parameter or static pressure \\
$Q_2$  & predictivity coefficient  \\
$q$  & number of clusters  \\
$r$  & specific gaz constant or correlation function  \\
$\bm{S}$  & matrix of the snapshots  \\
$s_i$  & quantity of interet at node $i$  \\
$T$ & temperature \\
$U$ & velocity \\
$w$  & weight of the Gaussian Mixture Model  \\
$X$  & horizontal coordinate along the chord  \\
$Y$  & vertical coordinate \\
$y$  & target value  \\

$\alpha$  & angle of attack  \\
$\Gamma$ & spatial domain \\
$\delta$  & Kronecker symbol  \\
$\epsilon$  & energy ratio  \\
$\theta$  & hyperparameters  \\
$\bm{\lambda}$  & eigenvalues matrix  \\
$\lambda$  & eigenvalues  \\
$\bm{\mu}$  & mean of the Gaussian Process  \\
$\rho$  & density  \\
$\bm{\Sigma}$  & covariance matrix  \\
$\sigma_0^2$  & prior covariance  \\
$\sigma$  & sigmoid function  \\

$\tau_w$ & wall shear stress \\
$\Phi$  & mixture coefficient  \\
$\bm{\phi}$  & proper orthogonal decomposition matrix  \\
$\bchi$  & input parameter  \\
${1}_{C}$  & hard splitting function  \\
\\

\noindent\textit{Subscripts} \\
$t$ & training \\
$p$ & prediction \\
$0$ & sea level \\
$\infty$ & freestream \\

\\
\noindent\textit{Superscripts} \\
$(k)$ & $k$-th component or element \\
$'$ & fluctuating part\\

\\
\noindent\textit{Operators} \\
\hspace{0.05 mm} \hphantom{l}$\widetilde{\cdotp}$ & surrogate model\\
\hspace{-0.2 mm} $\overbar{ \ \cdotp \ }$ & mean \\
\hspace{0.5 mm}$\left|\hspace{0.5 mm} \cdotp \right| $ & absolute value \\
$\lVert \hspace{0.5 mm} \cdotp \rVert_2$ & Euclidian norm \\
$( \cdotp , \cdotp)$ & canonical inner product \\ 
\end{longtable*}}

\section{Introduction} \label{introduction}

Overall aircraft design and optimization rely increasingly on numerical
simulations for structural, aerodynamics or even noise analysis. Particularly,
the computational fluid dynamics (CFD) plays a significant role in solving
Navier-Stokes equations, in order to predict vector-valued functions of specific
quantities of interest, such as wall pressure field. The equations are
discretized into algebraic systems that lead to prohibitive computational cost
for simulations with a high number of degrees of freedom. Moreover, the inflow
conditions may vary and form a multidimensional parameter space. Its full
exploration requires the computation of a very large number of expensive
simulations and becomes intractable. One of the main solution to overcome this
problem is the substitution of the high fidelity simulations by a mathematical
approximation much faster to be run, referred to as a surrogate model. It
represents an interesting trade-off between precision and computation time.
Furthermore, reducing the computational time of the exploration for high
fidelity CFD can open the way to multi-physics simulations using surrogate
models for the fluid parts.

The surrogate modeling of high-dimensional vector-valued functions is mainly
performed with a reduced-order approach, called reduced-order modeling (ROM).
Originally developed for the study of coherent structures in the turbulent
boundary layer~\cite{Aubry1988}, ROM methods have shown various applications
such as aeroelasticity~\cite{Amsallem2010}, optimal flow
control~\cite{Cordier2003}, turbulent flows~\cite{Berkooz1993, Osth2014} or
geophysics~\cite{Cstefuanescu2016}. Most ROM methods are applied to CFD problems
by approximating the high fidelity model as a linear combination of
low-dimensional basis vectors, weighted by purposely-tuned parameters. The
basis vectors characterize the main features of the system behavior. Proper
Orthogonal Decomposition (POD) is a particular and very popular method of
dimension reduction used very frequently for CFD problems. It computes the basis
vectors and the corresponding modal coefficients with an optimal least-square
approach from a given number of high-fidelity computations, also called
snapshots, at different state-parameters. The ROM coefficients are calculated
only for a finite and discrete number of input parameters whereas the surrogate
model is evaluated on a new parameter set. Thus a continuous representation of
the coefficients over the whole state-parameter space has to be provided in
order to build the final model, leading to two different approaches:
\begin{itemize}
	\item The intrusive ROM projects the governing equations into a set of basis
	functions of smaller dimension leading to a system of ordinary differential equations for
	the coefficients. The projection-based methods have the advantage to retain some
	of the physics from the governing equations and to give rigorous error bounds
	and error estimation~\cite{Veroy2005}. However, both stability and accuracy
	issues can occur~\cite{Xiao2016}. By their
	intrusive nature, these methods modify also the source code of the high fidelity model, leading
	to substantial modifications, if not impossible, when commercial software packages
	are used. Moreover, the reduced equations are solved on the whole domain
	and for all the conservatives variables even if the quantities of interest are
	evaluated on a subdomain and for a small number of variables. The high
	Reynolds \revision{number} turbulent
	flows~\cite{Protas2015,Lorenzi2016} give an example of challenging and active field of research for projection-based ROM.
	More information on intrusive ROM can be found in Benner
	\textit{et al.}~\cite{Benner2015}.

  \item The second method, the non-intrusive data fitting ROM, does not need any
  knowledge about the high-fidelity model, considered as a black box, allowing
  to deal with very complex physics. Instead of manipulating the governing
  equations, the value of the coefficients are predicted by methods of
  multidimensional data fitting such as Polynomial Regression, Radial Basis
  Function or Gaussian Process Regression~\cite{Forrester2008}.
  Non-intrusive ROM has been successfully applied in CFD, for instance in
  aero-icing problems~\cite{Fossati2013}, uncertainties quantification for urban
  flow~\cite{Margheri2016} or steady
  aerodynamics~\cite{BuiThanh2003parametricPOD,Braconnier2011,Mifsud2010}. One
  can note that hybrid methods mixing projection-based and non-intrusive
  approaches have been developed by solving an inverse problem, where the
  coefficients of the reduced-order governing equations are inferred using the
  output of the simulations~\cite{Peherstorfer2016}.
\end{itemize}
Here, only non-intrusive data fitting are considered and the
high-fidelity model is treated as a black-box.

Despite the usefulness of the energy-ranked POD and its extensive use, some
limitations have been observed~\cite{Graham1996}. Indeed, low energy
perturbations can be masked although they could be representative of a part
of the system behavior. Problems with bifurcations can have typical
characteristics, such as aerodynamic flows with varying inflow conditions leading to either
subsonic or transonic regime. In these cases, the classical method
computing the dominant modes in a single POD basis fails to produce accurate responses for
predictions in highly nonlinear region and not directly in the neighborhood of
the snapshots~\cite{Lucia2003,Iuliano2013}. The mix of the different physical
regimes in the POD basis vectors can explain this problem. Indeed, small errors
in the multivariate interpolation step can amplify POD modes associated with a
physical regime which does not exist for the considered prediction. For this reason,
approaches based on local reduced-order models have emerged in the literature by
considering only restrictions to the total amount of
snapshots~\cite{Eftang2010,Amsallem2012,Washabaugh2012,Franz2014, Kaiser2014,
Liem2015,Zhan2015}. This paper describes an original active local method, called
``\textit{Local Decomposition Method}'' (LDM), extending the classical
reduced-order modeling method using POD and data fit method to particular steady
problems with different physical regimes.

The method proposed here computes local subspaces of the state-parameter space
by combining a physical-based sensor with machine learning tools. The
physical-based sensor is a central element of the method to achieve proper
separations of the physical regimes. Indeed, the conversion of the vector-valued
output into \revision{a vector of physical-based features} gives the possibility
to cluster the snapshots into subsets with the same physical behavior. Thus the POD
basis vectors are more representative of the physics. A shock sensor is used for
the particular problems mixing subsonic and transonic conditions. It measures
the nonlinearities and sharp gradients of the flowfield. As the different
phenomena are no more mixed in the POD basis, building a local reduced-order
model on each of these subsets achieves a better consideration of the physical
regimes. The clustering of the snapshots provides also a greater flexibility to
the data fit model which can behave independently on each subgroup. As regards
the prediction of untried sets of parameters, a supervised learning algorithm
associates each region of the parameter space with a local reduced-order model
and its respective subspace, allowing to map the input parameter space to the
right physical regime. This last step is called input space decomposition.

Replacing the global POD basis with several local POD bases is a relatively recent
development for non-intrusive parametric reduced-order modeling and may seem
counterintuitive. Indeed, the compressive capability of the POD may be weakened
by increasing the number of bases, and the robustness of each data fit method
can decrease with the reduction of training samples due to the repartition of
these latter on the different models. On the contrary, the local models
enable a clear separation of the phenomena improving the prediction of the
surrogate model. It also increases the flexibility of the models of
data fitting. In this work, only Gaussian Process Regression (GPR) are
investigated as methods of data fitting. Thus the terms POD/data fitting and
POD/GPR are used indifferently to refer to the classical non intrusive ROM. 


This paper aims to present a local reduced-order model built with machine
learning tools and using a physical-based approach in order to address
parameter-dependent problems with either subsonic or transonic regime.
It is organized as follows: section~\ref{reduced_order_modeling} gives an
overview of the classical non-intrusive POD/data fitting approach. Then,
section~\ref{local_decomposition} introduces the LDM with its underlying
principles based on machine learning. Then, results from a one-dimensional
analytical case and two-dimensional transonic airfoil are presented in
section~\ref{applications}, demonstrating the capability of the LDM to deal with
different physical regimes including shock waves. Finally,
section~\ref{conclusion} provides a summary and the conclusions.

\section{Non-Intrusive POD/data fitting Reduced Order Modeling}
\label{reduced_order_modeling}
%
First, some notations are introduced. One considers a real vector-valued
function $\bm{f}$ representing the high-fidelity model defined from
$\mathds{R}^{p}$ to $\mathds{R}^{d}$, where $p$ is the number of parameters and
$d$ the dimension of the vector-valued quantity of interest. For example a CFD
code predicting the wall pressure field of an airfoil for different values of
Mach number and angle of attack defines an input domain with $p$ equal to $2$
and $d$ corresponding to the number of nodes representing the wall. Similarly,
$\widetilde{\bm{f}}$ represents the vector-valued surrogate model with the same
domain of definition from $\mathds{R}^{p}$ to $\mathds{R}^{d}$. The matrix of
the training input parameters is noted $\bchi_t = [\bchi_{t_1} \, \cdots \
\bchi_{t_n}]^T$  $\in$ $\mathds{R}^{n\times p}$ where $n$ is the number of
training samples, and $\bchi_{t_i}$  is the  $i$-th vector of the parameter set
which can be written with its components as $\bchi_{t_i} = [\chi_{t_i}^{(1)} \,
\cdots \ \chi_{t_i}^{(p)}]$. In the same way, the matrix of the  test samples,
also referred to as untried input parameters or merely the predictions, is noted
$\bchi_p = [\bchi_{p_1} \, \cdots \ \bchi_{p_m}]^T$  $\in$ $\mathds{R}^{m \times p}$
with $m$ the number of predictions. $\bm{S}_i$ designates the vector of the
$i$-th snapshot such that $\bm{S}_i = \bm{f}(\bchi_{t_i})$ $\in$ $\mathds{R}^{d}$
and $\bm{S}$ defines the matrix of the training snapshots $\bm{S} = [\bm{S}_1 \,
\cdots \ \bm{S}_n]^T$ $\in$ $\mathds{R}^{n\times d}$. The terms
$\overbar{\bm{S}}$ and $\overbar{\bm{f}}$ are used interchangeably to refer to
the mean snapshot $\overbar{\bm{S}} = \overbar{\bm{f}} =
\frac{1}{n}\sum_{k=1}^{n} \bm{S}_k \in \mathds{R}^{d}$.
The fluctuating part $\bm{S}'$ of the snapshots is defined by the snapshots
matrix where the mean snapshot has been removed such that $\bm{S}' = [\bm{S}_1 -
\overbar{\bm{S}} \, \cdots \ \bm{S}_n - \overbar{\bm{S}}]^T$.

\begin{figure}[!h]
\centering
\includegraphics[width=0.5\textwidth]{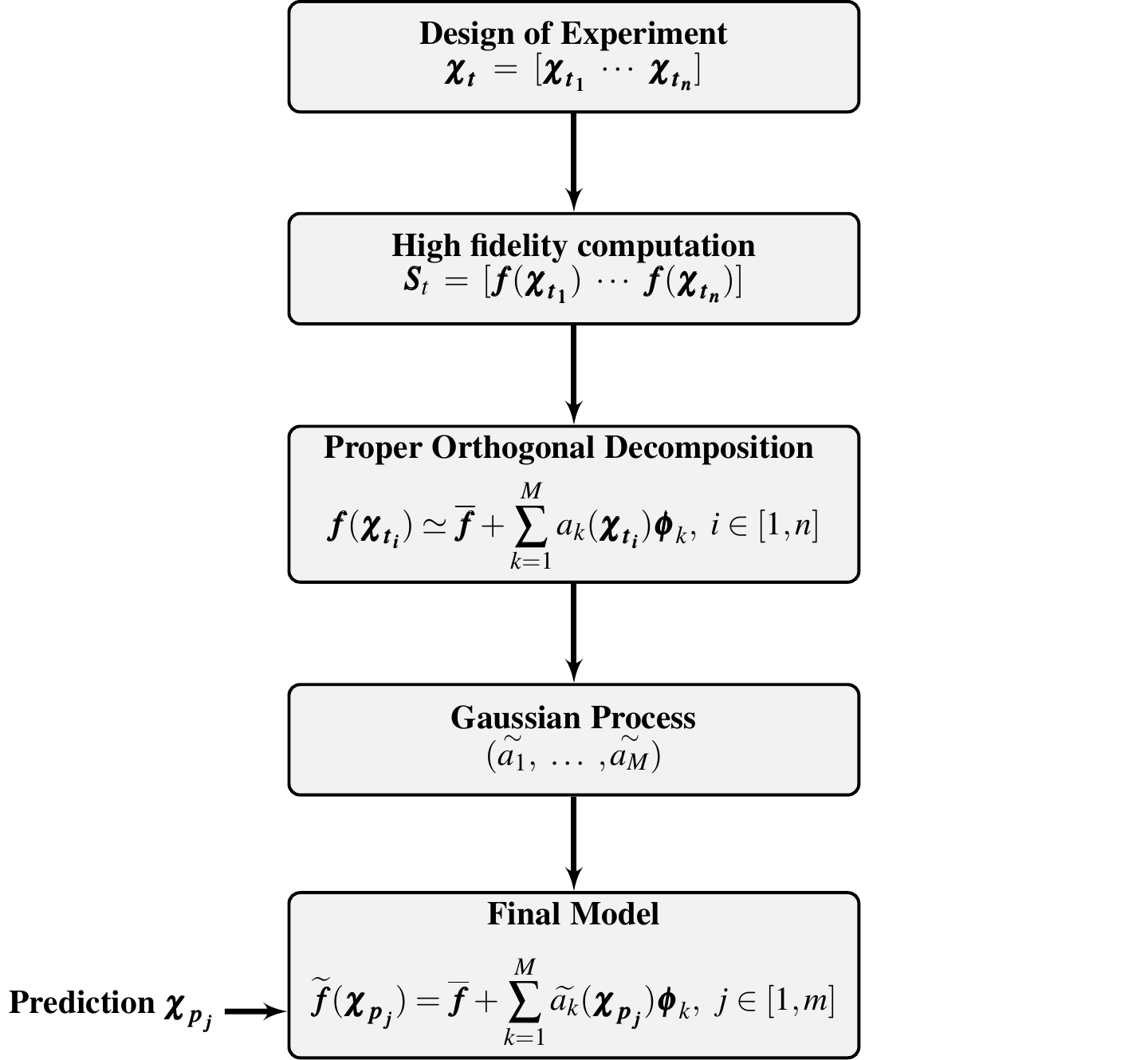}
\caption{Flowchart of the POD/GPR method.}
\label{fig:jpod_workflow}
\end{figure}

The non-intrusive POD/data-fit reduced order modeling is made of three steps:
the generation of the initial training samples, the POD dimension reduction and
the pseudo-continuous representation of the POD coefficients. All of them are
described in the following parts. The whole process of the method is depicted
in~\figurename{\ref{fig:jpod_workflow}}. One can note that the POD/GPR method is
the common basis of the further developments presented in this paper and serves
as a reference for the method assessment performed in
section~\ref{applications}.
\subsection{Sampling plans}
The purpose of the surrogate model is to simulate the input/ouput behavior over
the domain of variation of the parameters based on a limited number of
high-fidelity evaluations. Correct predictions are required not only for the
design points but also at all off-design conditions. For this reason, the
limited number of snapshots should be optimally placed in order to build a model
capturing the maximum amount of information about the physics over the parameter
space. An inappropriate repartition of the input parameters could lead to a
surrogate model with large discrepancies. To the extent possible, the number of
snapshots will be limited to the rule of thumb $10p$ studied by Loeppky
\textit{et al.}~\cite{Loeppky2009}, where $p$ is the number of parameters.

Contrary to the projection-based method, the non-intrusive POD/GPR approach
does not have access to prior information on the system given by the
coefficients of the governing equations~\cite{Paul2015}.
The critical issue of the choice of the a priori training snapshots is called
Design Of Experiment (DOE). As explained previously, the main goal of the DOE is
to generate well-distributed samples in the parameter space to give sufficient
information to the learning process. DOE methods have been widely studied in the
literature providing many techniques for experimental parametric studies and
computer experiments. One can cite for example random and orthogonal array
methods with Monte Carlo and Latin Hypercube Sampling~\cite{Mckay1979},
geometrical approaches such as centroidal Voronoi
tessellations~\cite{Quiang1999} or low-discrepancy sampling techniques like
Halton, Sobol or Faure sequences~\cite{Kalos2008}. In the present work,
low-discrepancy sequences have been adopted due to their iterative design. Indeed, the number of samples
can be extended on purpose and high-density regions can be easily defined while
keeping the space-filling properties. Both properties are very interesting
features for active learning. The other deterministic methods require a pre-set
number of samples and cannot be extended without losing a part of their
space-filling property.
\subsection{Proper Orthogonal Decomposition}
The POD is an efficient technique of dimension reduction based on spectral
decomposition for high dimensional, multivariate and nonlinear data set. A
\revision{wide} range of applications can be found in the literature such as
human face characterization~\cite{Kirby1990}, data compression~\cite{Andrews1967} or optimal control~\cite{Lombardi2011}. The
POD term was first introduced in 1967~\cite{Lumley1967} to study dominant
turbulent eddies, also called Coherent Structures.
POD is also known as Karhunen-Lo\`eve Decomposition, Hotelling Analysis or
Principal Component Analysis, in other fields of application. Among all the
possible linear decompositions of the high-fidelity function, the POD method
minimizes in a least square sense the residual of the projection of the
high-fidelity model, yielding an optimal basis in term of the
representativeness of the data~\cite{Cordier2007}.
The least square problem is equivalent to a maximization
problem~\cite{Berkooz1993, Cordier2007}. Introducing the canonical inner product
$(\cdotp,\cdotp)$ on $\mathds{R}^{d}$, the POD basis is the solution of
the following formulation:
\begin{equation}
\begin{aligned}
\max\limits_{\bm{\phi_1}, \ldots, \bm{\phi_n} } \sum_{i=1}^n \sum_{j=1}^n
|(\bm{S}'_i,\bm{\phi}_j)|^2  \\
\text{subject to } (\bm{\phi_i},\bm{\phi_j}) =  \delta_{i,j} 
\end{aligned}
 \label{eq:pod_max_pb}
\end{equation}
where $\bm{\phi}_i\in \mathds{R}^{d} $ is the $i$-th vector basis, $\bm{S}_i'$
the fluctuating part of the $i$-th snapshot and $\delta_{i,j}$ the Kronecker
symbol satisfying $\delta_{i,j} = 1$ for i = j and $\delta_{i,j} = 0$ otherwise.
The first mode is very close to the mean value of the snapshots for reasonable
variations in the data. For this reason, the POD is performed on the fluctuating
quantity $\bm{S}'$. The matrix of the vector basis $\bm{\phi} \in \mathds{R}^{n
\times d}$ is introduced such that $\bm{\phi} = [\bm{\phi}_1 \ \cdots \
\bm{\phi}_n]^T$.

The method of snapshots proposed by Sirovich~\cite{Sirovich1986} is employed to
solve the maximization problem in Eq.~\ref{eq:pod_max_pb} and leads
to an eigenvalues problem:
\begin{equation}
\frac{1}{n} \bm{S}'\bm{S}'^T =  \bm{\phi} \bm{\lambda} \bm{\phi}^T
\label{eq:pod_discre_classical_eigenproblem}
\end{equation}
where $\bm{\lambda} \in \mathds{R}^{d \times d}$ is the diagonal matrix of the
eigenvalues associated to the matrix of the eigenvectors $\bm{\phi}$. The
eigenvalues problem can be solved either by an eigen-decomposition or a
Singular Values Decomposition (SVD). The latter gives a better precision for
the smaller eigenvalues and provides an iterative approach well-fitted for
resampling. One can note that the self-adjoint operator property of
$\bm{S}'\bm{S}'^T$ ensures that the computed POD modes form a complete
orthonormal set built as $\{\bm{\phi}_1,\ \ldots,\ \bm{\phi}_n \}$, on which
the high-fidelity model is decomposed:
\begin{equation}
 \bm{f}(\bchi_{t_i})= \overbar{\bm{f}} + \sum_{k=1}^{n}
 a_{k}(\bm{\bchi}_{t_i})\bm{\phi}_{k}, \ \forall i \in [1, n]
 \label{eq:basis_decomposition}
\end{equation}
where $a_k(\bm{\bchi}_{t_i}) \in \mathds{R}$ is the reduced coordinate
associated with the $k$-th POD mode $\bm{\phi}_k \in \mathds{R}^d$. All the reduced
coordinates are computed using the orthonormality property of the POD basis and are
expressed as:
\begin{equation}
 a_k(\bchi_{t_i})= (\bm{S}'_i, \bm{\phi}_k) 
 \label{eq:reduced_coordinate}
\end{equation}
Since the POD is optimal in term of energy and provides an energy-ranked basis,
only a small number of the most energetic POD modes can be retained in order to
reduce the dimension of the system. The smallest eigenvalues are neglected
leading to a truncation of the basis. This heuristic criterion can be written
more formally as a minimum ratio of the captured energy, which means finding the
number $M$ of kept basis vectors such that for a given amount of energy ratio
$\epsilon$:
\begin{equation}
 \frac{\sum\limits_{k=1}^{M} \lambda_k} {\sum\limits_{j=1}^{n} \lambda_j} >
 \epsilon
\end{equation}
A classic ratio of energy present in the literature is 0.99~\cite{Cordier2003,
Sirovich1986}. Once $M$ has been set, the truncated linear combination of the
eigenfunctions gives the approximation of the high-fidelity model:
\begin{equation} 
 \bm{f}(\bchi_{t_i})\simeq \overbar{\bm{f}} + \sum_{k=1}^{M}
a_k(\bchi_{t_i}) \ \bm{\phi}_{k} \ \ \forall i \in [1,n]
 \label{eq:pod_final}
\end{equation}

\subsection{Pseudo-continuous representation with Gaussian Process Regression}
The pseudo-continuous representation is the next step of the POD/GPR surrogate
model. The reduced coordinates $a_k$ have been computed at a small number of
training parameters whereas the 
analysis of the high-fidelity model for various inflow conditions requires a
continuous evaluation over the input parameter space.
Thus, the values of the reduced coordinates at untried parameter combinations
are estimated with a data-fit method. The most popular methods in the surrogate
modeling literature are formed of polynomial regression, Radial Basis Function
or Gaussian Process Regression (GPR)~\cite{Forrester2008}. A particular emphasis
is given to this latter method. It has been employed in this paper due to its
capability to deal with nonlinear problems, its high flexibility and the
provided error estimation of the predictor. GPR is also called ''Kriging'' and
has been first applied in geostatistics~\cite{Krige}. The short overview of the
GPR in this paper is introduced following the formalism of
Rasmussen~\cite{Rasmussen2005} and is directly applied to the continuous
representation of the reduced coordinates. They are assumed to follow a
Gaussian process, which is outlined by a collection of random variables having a
joint Gaussian distribution of mean $\bm{\mu}$ and covariance matrix
$\bm{\Sigma}$. If the reduced coordinate $a_k$ follows a Gaussian process, it
reads:
\begin{equation} 
\bm{A}_t^{(k)} \sim \mathcal{N}(\bm{\mu}^{(k)}, \bm{\Sigma}^{(k)})
\end{equation}
with $\bm{A}_t^{(k)}=[a_k(\bchi_{t_1}) \ \cdots \ a_k(\bchi_{t_n})]^T$ defining the
matrix of the $k$-th reduced coordinate at training parameters, and
$\bm{A}_p^{(k)}=[a_k(\bchi_{p_1}) \ \cdots \ a_k(\bchi_{p_m})]^T$ the predictions at
the unknown combinations of parameters. In the interests of simplifying notation
and analysis, the index $k$ is removed and becomes implicit. The joint
distribution of the reduced coordinates at training and unknown parameters is given by:
\begin{equation}
\begin{bmatrix}
\bm{A}_t \\
\bm{A}_p \\
\end{bmatrix}
 \sim \mathcal{N} \left(
\begin{bmatrix}
\bm{\mu}_t \\
\bm{\mu}_p \\ 
\end{bmatrix}
 ,
 \begin{bmatrix}
\bm{\Sigma}_{tt} & \bm{\Sigma}_{tp} \\
\bm{\Sigma}_{pt} & \bm{\Sigma}_{pp} \\
\end{bmatrix}\right)
\end{equation}
with $\bm{\Sigma_{tp}}$ the covariance matrix between $\bchi_t$ and $\bchi_p$.
The central issue of the GPR remains to be addressed, namely how to determine
the means and the covariance between the inputs. A classical stationarity
assumption is that the correlation depends only on the magnitude of the
Euclidean distance between the two input parameters but not
on the values themselves, such that the $i$, $j$-th element of the covariance
matrix is given by:
\begin{equation}
[\bm{\Sigma_{tp}}]_{i,j} = \sigma_0^2 r\Big(\lVert \bchi_{t_i} - \
\bchi_{p_j} \rVert_2 \Big) \ \ \forall i,j \in [1, n]\times [1, m]
\end{equation}
with $\sigma_0^2$ the prior covariance corresponding to the level of uncertainty
for predictions far from the training data and $r$ the correlation function.
The latter is usually monotonically decreasing with $r(0)=1$. A wide range
of functions have been proposed to model the relation between the covariance and
the input distance, such as Radial Basis Function, Mattern or periodic
regression function~\cite{Rasmussen2005}. The anisotropic Radial Basis Function,
chosen for our problem, is one the most classical model of the regression
functions due to its smoothness, infinite differentiability and analytical
derivability. Its expression introduces the hyperparameter $\theta_k \in
\mathds{R}^+$ and is given by:
\begin{equation}
 r\Big(\lVert \bchi_{t_i} - \
\bchi_{p_j} \rVert_2 \Big) = \prod_{k=1}^{p} \exp\left(\frac{- \lVert
 \chi_{t_i}^{(k)} - \chi_{p_j}^{(k)} \rVert_2 } {2\theta_k} \right) \ \ \forall
 i,j \in [1, n]\times [1, m] 
\label{toto}  
\end{equation} 
The hyperparameter $\theta_k$ defines the way the data are explained by the
component $k$ of the input parameters. Small values means the correlation is
high between the inputs and the model is very sensitive to the dimension $k$.
On the other hand, large values of $\theta_k$ illustrates a model slowly varying
with the data.

The final form of the predictor is derived using the conditional distribution of
$\bm{A}_u$ given $\bm{A}_t$, also called a posteriori distribution which is
still Gaussian and written as:
\begin{equation}
p(\bm{A}_p|\bm{A}_t) \sim \mathcal{N}(\bm{\mu}_p + \bm{\Sigma}_{pt}
\bm{\Sigma}_{tt}^{-1}(\bm{A}_t - \bm{\mu}_t), \ \bm{\Sigma}_{pp} -
\bm{\Sigma}_{pt} \bm{\Sigma}_{tt}^{-1} \bm{\Sigma}_{tu})
\end{equation}
The mean of the distribution gives the final value of the predictions at the
untried set of parameters. Regarding the variance, it provides an estimate of
the possible range taken by the prediction which can also be seen as the
mean-square error. The latter has the interesting feature not to be dependent on
the value of the output but only to the input parameters. Both $\theta_k$ and
$\sigma^2_0$ remain to be determined in order to obtain the final prediction.
They are computed during the training phase of the GPR, most of the time by a
Maximum Likelihood Estimation (MLE) approach or a leave-one-out
method~\cite{Rasmussen2005}. Martin \textit{et al.}~\cite{Martin2005} has shown
the MLE works better than leave-one-out in general. For this reason, MLE is
applied in the GPR/POD method, solving numerically the nonlinear maximization
problem of the log-likelihood:
\begin{equation}
\log(p(\bm{A}_t | \bm{\theta})) = - \frac{1}{2} \bm{A}_t^T
\bm{\Sigma}_{tt}^{-1}\bm{A}_t - \frac{1}{2} \log|\bm{\Sigma}_{tt}| -
\frac{n}{2}\log(2\pi)
\end{equation}
where $|\cdotp|$ denotes the determinant operator.
The partial derivatives of the marginal likelihood with
regards to the hyperparameters can be analytically derived.
Thus, it is possible to use a gradient-based optimization
algorithm in order to numerically find a local solution to the MLE problem. The
Limited-memory Broyden-Fletcher-Goldfarb-Shanno Bounded (L-BFGS-B)
algorithm~\cite{Byrd-1995} is employed in this paper to determine the
hyperparameters. This popular quasi-Newton method handles simple
bound constraints and is coupled with random restarts to avoid local maximum of
bad quality.

By assuming the POD basis vectors are invariant with respect to the input
parameters, the final surrogate model predicts the quantity of interest at the
$j$-th untried input parameter $\bchi_{p_j}$ such that:
\begin{equation}
 \widetilde{\bm{f}}(\bchi_{p_j}) = \overbar{\bm{f}} + \sum_{k=1}^{M}
 \widetilde{a_k}(\bchi_{p_j})\bm{\phi}_{k}, \ \forall j \in [1,m]
 \label{eq:gpr_final}
\end{equation} 
with $\widetilde{a_k}(\bchi_{p_j}) = [\bm{\mu}_p^{(k)} +
\bm{\Sigma}_{p_jt}^{(k)} {\bm{\Sigma}_{tt}^{(k)}}^{-1}(\bm{A}_t^{(k)} -
\bm{\mu}_t^{(k)})]_{j}$ the approximation of the weighting coefficient $a_k$
over the parameter space. The a priori mean $\bm{\mu}_p^{(k)}$ is usually
considered equal to zero as the training data have been standardized with zero
mean. One can note that different versions of GPR or Kriging can be used, such
as Bayesian Kriging \cite{Fossati2015}. Here, the Python library
scikit-learn~\cite{scikit-learn} is employed to generate the GPR models.

\section{Local Decomposition Method} \label{local_decomposition}
\begin{figure}[!h]
\centering 
\includegraphics[width=0.42\textwidth]{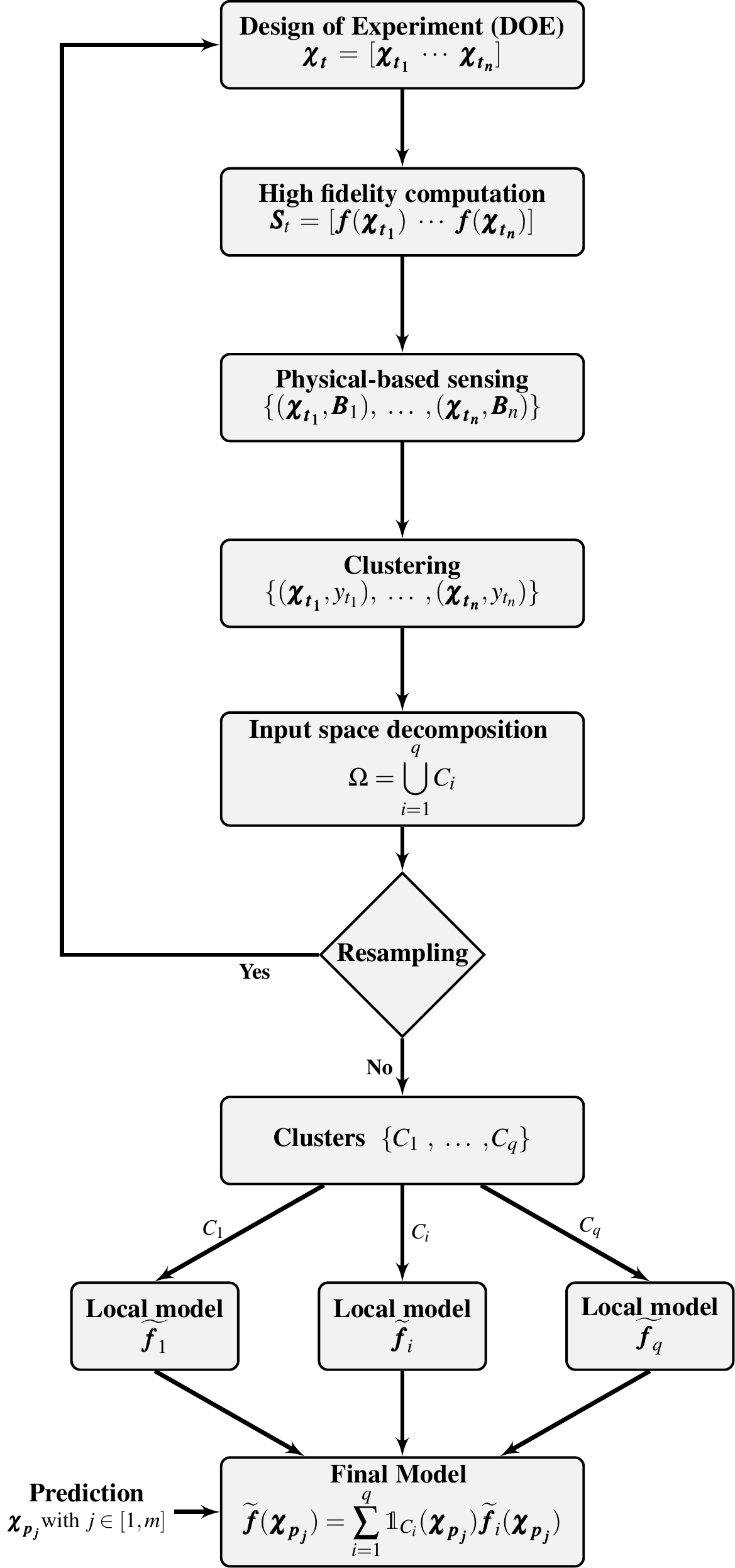}
\caption{Flowchart of the LDM.}
\label{fig:ldm_workflow}
\end{figure}

As explained in the introduction, the LDM proposed in this paper, illustrated
in~\figurename{\ref{fig:ldm_workflow}}, extends the classical POD/GPR
reduced-order modeling by employing a local approach, inspired by the mixture of
experts~\cite{Liem2015} and dynamic local reduced-order
modeling~\cite{Amsallem2012}. Instead of a unique global POD basis, several
local bases are computed using machine learning tools yielding to more flexible
behaviors bringing out a precise delimitation of the physical regimes. One can
note that a comparable approach has been used for aero-icing
certification~\cite{Zhan2015}. \revision{The specificity of the presented
method includes the introduction of a feature extraction with a shock sensor, a
novel resampling strategy and the application to a aerodynamics case.} \revision{A
shock sensor computes new features in order to ease the clustering of the
snapshots.} \revision{Moreover, an active re-sampling is carried out by
identifying the subspace with the highest entropy. Adding extract snapshots in
these specific subspaces will minimize the redundancy of the sampling.}

\subsection{Coupling machine learning tools with a physical sensor}
Let us introduce basic machine learning vocabulary. Learning problems can be
divided into two distinctive categories: \textit{supervised} and
\textit{unsupervised}. Here, the machine learning library
\textit{scikit-learn}~\cite{scikit-learn} is employed in the in-house JPOD code
to perform both supervised and unsupervised learning. In the context of
supervised learning, some input variables have an influence on one or more
outputs. This set of inputs and outputs forms a learning base, and the
supervised learning simulates the input/output behavior using the learning base.
The final goal is to predict the values of the outputs for untried inputs. The
nature of the output subdivides the supervised learning into two subcategories:
the classification, dealing with categorical input variables, and the regression
which is applied on real and continuous input variables. The GPR is an example
of regression. As regards the unsupervised learning, the training set consists
only of the input vectors without any corresponding outputs. Thus the purpose of
the unsupervised learning is to identify underlying structures hidden in the
parameter space but the accuracy of the algorithm cannot be defined by any
objective function.

\subsubsection{Physical-based shock sensor to detect flow regimes}
The choice of the quantity characterizing the physical regimes, on which the
clustering is performed, is a question of central importance impacting the
quality of the classification. Usually, the unsupervised learning clusters
directly the quantity of interest into groups with patterns of small
differences~\cite{Amsallem2012,Washabaugh2012,Zhan2015}. However, the aim of
the clustering in this paper is the physical regime separation and the previous
approach can lead to classification error. Indeed, two fields of the quantity of
interest can have large differences even tough they belong to the same
physical regime. A classical method fails to separate them accurately. For this
reason, this section proposes another method to perform the clustering. Based on a physical approach, a
mathematical transformation converts the quantity of interest into a sensor of
the physical regime. The main goal is to sharply quantify the physical regime to
ease the clustering of the snapshots.

The application of this paper involves external aerodynamics with
subsonic and transonic regimes, characterized by shock waves. A straightforward
idea is to consider a shock sensor that is able to detect large changes in the
variation of the quantity of interest, such as Jameson's Shock
sensor~\cite{Jameson1981}. It is related to the second order
derivative of the pressure.
However, the quantity of interest is not limited to pressure signals and a more
general expression providing an estimation of the second derivative is
introduced as:
\begin{equation}
\nu_i = \frac{| s_{i+1} - 2s_i + s_{i+1}|}{\epsilon_0 + |s_{i-1}|
+2|s_i|+|s_{i+1}|}, \ \forall i \in [2,d-1]
\end{equation}  
where $\nu_i$ is the generalized sensor, $\epsilon_0$ a constant avoiding
division by $0$ and $s_i$ the quantity of interest. The dimension of the
generalized sensor can be reduced by POD, such as the $k$-th POD
basis vector is associated to the reduced coordinates $b_k$. The latter can
be interpreted as the representative quantity of the physical regime for a
given snapshot. They are grouped in the vector $\bm{B}_i$ defining the matrix $\bm{B} = [\bm{B}_1 \
\cdots \ \bm{B}_n]^T.$

\subsubsection{Clustering of the  shock sensors by Gaussian
Mixture Model}
The problem of identifying the inherent groupings in the input
data refers to unsupervised classification, specifically clustering. K-means and
Gaussian Mixture Model (GMM) are two well-known examples of algorithms
classically employed to achieve clustering. This section puts a particular emphasis on GMM
algorithm~\cite{Mclachlan1988}, where the main features are described.

Let us assume the set $\{\bm{B}_1 ,\ \ldots \ ,  \bm{B}_n\}$, characterizing the
physical regimes, comes from $q$ clusters $C_1, \ \ldots ,\ C_q$. Each cluster
$C_k$ follows a probability distribution of parameter $\theta_k$ and proportion $w_k$, regrouped for all the clusters in a mixture
parameter $\Phi = [w_1 \ \cdots \  w_q \ \theta_1 \ \cdots \  \theta_q]$. GMM
consists in modeling $\bm{B}$ with a mixture distribution of multivariate normal
distributions $g$. Each one is associated with the cluster $C_k$ such that the
probability density function of $\bm{B}_i$ is given by:
\begin{equation}
p(\bm{B}_i| \Phi) = \sum_{k=1}^q w_k \, g(\bm{B}_i| \theta_k), \ \forall i \in
[1, n]
\label{eq:gmm}
\end{equation}
the mixture weights $w_k$ represent the probability that the observation
comes from the $k$-th Gaussian distribution and $\theta_k$ gives the mean and
the covariance of the multivariate normal distribution $g$.

These mixture parameters regrouped in $\Phi$ are estimated iteratively using an
Expectation Maximization algorithm. The probability $p$ of $B_i$ belonging to
the cluster $k$ can be expressed with Bayes' theorem:
\begin{equation}
p(\bm{B}_i \in C_k|\bm{B}_i) = \frac{p(\bm{B}_i|\bm{B}_i \in C_k) \,
p(\bm{B}_i \in C_k)} {P(\bm{B}_i)} = \frac{w_k \, g(\bm{B}_i|
\theta_k)}{  \sum\limits_{l=1}^q w_l  \, g(\bm{B}_i| \theta_l)} , \ \forall i \in
[1, n]
\end{equation}
The cluster of each quantity $\bm{B}_i$ can be determined using the previous
probability expression. The training set is built by applying a hard
splitting such that:
\begin{equation}
\{ (\bchi_{t_1}, y_{t_1}), \ \ldots \ , (\bchi_{t_n}, y_{t_n}) \}  \text{ with
} y_{t_i} = \left\{C_k \ | \ \max\limits_{k \in [0, q]} 
p(\bm{B}_i \in C_k|\bm{B}_i) \right\} , \ \forall i \in [1, \ n]
\end{equation}
with $y_{t_i}$ the target variable of the cluster. The training set is used to
train a supervised learning algorithm, described in the following section, in
order to link the input parameter of each quantity $\bm{B}_i$ with a class of
physical regime.
\subsubsection{Input space decomposition using Gaussian Process Classification} 
The decomposition of the input space into subspaces where a single physical
regime drives the flow can be interpreted as a supervised classification
problem. Indeed, the training set is provided by the clustering phase and trains
an algorithm assigning the $k$-th untried input parameters $\bchi_{p_j}$
($j\in[1,m]$) to the categorical variable $y_{p_j}$ which can
take the different values of the $q$ classes $C_1, ..., C_q$. The Gaussian
Process Classification (GPC) is a classical method to deal with classification. The principal steps of the method for
two-class problem are outlined in this section. The interested readers can refer
to Rasmussen \textit{et al.} and Bishop~\cite{Rasmussen2005, Bishop2006} for the
straightforward generalization to $K$ classes.

As the nature of the classification outputs is discrete, it clearly differs from
the regression problems outlined previously with GPR. The main idea is to
transform the output of a Gaussian process defined on the real axis into a
probability lying in the interval $[0,1]$ using a nonlinear activation function.
A latent function ${l}$ defined on the input parameter is introduced and we
denote the latent vector by $\bm{l}$ such that:
\begin{equation}
  \bm{l} = [l_{t_1}, \ldots, l_{t_n}] \ \text{with} \ l_{t_i} =
  l(\bchi_{t_i})  \ \text{and} \ l_{p_j} =
  l(\bchi_{p_j}), \ \ \forall i, j \ \in [1,n] \times [1,m] 
\end{equation}
This function aims to provide a more convenient and tractable formulation of the
model~\cite{Rasmussen2005} and will be removed by integration. A Gaussian
process prior with a zero mean and a covariance matrix $\bm{\Sigma}_{l}$ is
placed on the joint distribution of the latent function ${l}$: 
\begin{equation}
 \begin{bmatrix}
 \bm{l} \\
l_{p_j} \\
\end{bmatrix}
 \sim \mathcal{N}(\bm{0}, \bm{\Sigma}_{l})
\end{equation}
As regards the two-class problem with $C_0$ and $C_1$, the probabilistic
prediction is directly computed by $p(y_{p_j} = C_0 | \bchi_t, y_t, \bchi_{p_j})$
because $p(y_{p_j} = C_1 | \bchi_t, y_t, \bchi_{p_j})$ is given by $1-p(y_{p_j} = C_0 |
\bchi_t, y_t, \bchi_{p_j})$. The conditioning on the input variables is
intentionally let implicit. The probabilistic prediction is given by:
\begin{equation}
    p( y_{p_j} = C_0 |y_t) = \int p(y_{p_j} = C_0 | l_{p_j})p(l_{p_j} |y_t) dl_{p_j}
\label{eq:untractable}
\end{equation} 
where $p(y_{p_j} = C_0 | l_{p_j})= \sigma(l_{p_j})$ with $\sigma$ the nonlinear
activation function defined by the sigmoid function:
\begin{equation}
\sigma(x) = \frac{1}{1+ e^{-x}}
 \label{eq:sigmoid}
\end{equation}
The integral expressed in Eq.~\ref{eq:untractable} is analytically
intractable due to the non-Gaussian likelihood of
$p(l_{p_j}|y_t)$~\cite{Rasmussen2005}. The expansion of the latter with the sum
rule, product rule and Baye's theorem gives:
\begin{equation}
   p(l_{p_j}|y_t) = \int p(\bm{l} | y_t)p(l_{p_j}|\bm{l}) d\bm{l}
\end{equation} 
where $p(l_{p_j}|\bm{l})$ is Gaussian. The non-Gaussian probability
$p(\bm{l}|y_t)$ requires specific approximations, such as variational inference,
expectation propagation or Laplace approximation~\cite{Bishop2006}. Finally, the
hyperparameters of the covariance matrix $\bm{\Sigma}_{l}$ need to be
determined, for example with the maximization of the log-likelihood, which also
required to use the Laplace approximation due to non-Gaussian terms.


\subsection{An entropy-based active resampling}
Several methods have been coupled with surrogate models to generate an active
reduced-order model: local methods and subspace methods form the two main
different approaches. Local methods look for particular points which could
improve the accuracy of the model. One can cite for example leave-one-out
cross-validation~\cite{Fossati2015} testing the sensibility of the surrogate
model to each training sample. The more critical sample for the surrogate
defines a neighborhood in which an extra snapshot is added. Another example of a
local method, the Maximum Mean-Squared Error, uses the posterior estimation of
variance from the GPR and adds the sample with the maximum value to the training
set. Multi-fidelity can also be used to perform active infill
sampling~\cite{Benamara2016}. Another type of method based on a subspace
approach identifies particular low-dimensional structures in the input
parameters where the quantity of interest shows a significant variability. One
can cite for example active subspaces~\cite{Constantine2013} or sensibility
analysis~\cite{Saltelli2008}.

The original strategy proposed in this paper is based on a subspace approach
and aims at taking advantage of the clustering. Indeed, the input space
decomposition has provided subgroups of smaller dimensions among which some subspaces of interest
can be selected to perform the resampling. One proposes to use a criterion based
on the compressibility of the information, coming from an analysis of the POD
eigenvalues, in order to identify these relevant structures. As explained in the
last section, the POD eigenvalue represents the relative information contained
by the modes. The global entropy $H$ measures the redundancy of this information
and is introduced
as~\cite{Cordier2007}:
\begin{equation}
H = - \frac{1}{\revision{\log} n}\sum_{k=1}^{n}p_k \log(p_k) \text{ with } p_k
= \frac{\lambda_k}{\sum\limits_{i=1}^{n}\lambda_i}
\end{equation}
If the entropy goes to zero, there is only
one nonzero singular value. The data are compressed into a unique mode. On the
other hand, the entropy is equal to one if all the information is distributed
among the modes, meaning that no compression is possible. Between these two
extreme values, the entropy increases with the number of fundamental modes. The
active resampling of the LDM assumes that the entropy and the nonlinear
structures of the system are directly correlated. Thus, the probability to find
new modes with a non-negligible amount of energy is expected to be greater for
the cluster with the highest value of entropy than for any other cluster.

\subsection{Recombination in a global model by hard-splitting}
The final recombination step consists in assembling the local reduced-order
models in a single composite global model. Starting from the $q$ clusters, a
simple weighted sum is calculated using a 'hard' split:
\begin{equation}
 \widetilde{\bm{f}}(\bchi_{p_j}) = \sum_{i=1}^{q}
 \mathds{1}_{C_i}(\bchi_{p_j}) \widetilde{\bm{f}_i}(\bchi_{p_j}),
  \ \forall j \in [1,m]
 \label{eq:ldm_final}
\end{equation}
where
\begin{equation}
 \mathds{1}_{C_j}(\bchi)=\begin{cases} 
               1\text{ if } j = \argmax\limits_{i \in [1, q]} P(\bchi
               \in C_i)
               \\
               0 \text{ else }
            \end{cases}
\end{equation}
and $\widetilde{\bm{f}_i}$ refers to the classical POD/GPR model built on the
$i$-th cluster. This sum provides a continuous but not differentiable prediction
of all the input space, leading to a global model. The differentiable predictions require to use differentiable weighting functions, which is not the case for $\mathds{1}$. One can cite for example soft clustering~\cite{Bettebghor2011},
substituting $\mathds{1}_{C_j}(\bchi)$ directly by $P(\bchi \in C_i)$. However, it mixes
several physical regimes, leading potentially to unphysical predictions and can
amplify the extrapolation of the reduced coordinates near the boundary decision.
For these reasons, a hard-split approach has been selected.

The decision boundary in the input space parameters poses another problem.
The classification is very prone to errors in this region.
Indeed, the localization of the decision boundaries is subject to local
variations, such as the choice of the supervised learning method
(model-based, local methods, \ldots) or the location of the training samples.
Moreover, the reduced coordinates can be in extrapolation in this region. To
overcome these problems, the classical model is employed for the predicted
points near the interface.


\section{Numerical results} \label{applications}

\subsection{Error measurements}
The classical method and the proposed LDM are assessed in this part with the
RAE2822 airfoil~\cite{Benamara2016,Iuliano2013}. It may be noted that an
additional study has been performed on the Burgers' problem. The resultats are
shown in the supplemental material burgers\_am.pdf. The flow around the
two-dimensional transonic airfoil is computed with a Navier-Stokes solver, involving a turbulence model. It is a challenging
application in term of surrogate modeling with high discontinuities due to the
appearance of shocks. A three-dimensional input space is considered.

Particular attention is paid to the process of the input space
decomposition and to the comparison of both classical approach and LDM in term
of accuracy. Several quantities are introduced to measure the accuracy.
The so-called predictivity coefficient $Q_2$ gives the ratio of the output
variance which is explained by the metamodel. It can be interpreted as the
classical coefficient of determination of the linear regression applied to a
test sample~\cite{Iooss2010}. The more the value is close to 1, the higher
variance is explained by the model. It is expressed as:
\begin{equation}
Q_2^{(i)} = 1 - \frac{\sum\limits_{j=1}^{m} [ f^{(i)}(\bchi_{p_j}) -
\widetilde{f}^{(i)}(\bchi_{p_j})]^2}{\sum\limits_{j=1}^{m} [ \overbar{f}^{(i)}-
f^{(i)}(\bchi_{p_j})]^2}
\label{eq:q2}
\end{equation}
The Root Mean Square Error (RMSE) and the Normalized Root Mean Square Error
(NRMSE) are also introduced: 
\begin{equation}
RMSE^{(i)} = \sqrt{\frac{1}{m}\sum\limits_{j=1}^{m} [
f^{(i)}(\bchi_{p_j}) - \widetilde{f}^{(i)}(\bchi_{p_j}) ]^2}
\label{eq:rmse}
\end{equation}
\begin{equation}
NRMSE^{(i)} = \frac{RMSE^{(i)}}{f_{max} - f_{min}}
\label{eq:nrmse}
\end{equation}
where $f_{max}$ and $f_{min}$ refer respectively to the maximum and minimum
value of the function to predict $\bm{f}$. One can note that the $Q_2^{(i)}$ and
$NRMSE^{(i)}$ are computed at a given index $i$ of the spatial domain $\Gamma$.
The global measures are provided by averaging the quantities over this domain.
The global quantities are referred to $\langle Q_2 \rangle_{\Gamma}$ and
$\langle NRMSE \rangle_{\Gamma}$, with $\langle . \rangle_{\Gamma}$ the spatial
average operator. However, in order to provide also a statistical error
analysis, the averaged normalized error $E_i$ is introduced. It corresponds to the absolute error between the exact value and the
prediction, normalized by the range of variation, at snapshot level $j$:
\begin{equation}
E_j = \frac{ \langle |\bm{f}(\bchi_{p_j}) - \widetilde{\bm{f}}(\bchi_{p_j})
|\rangle_{\Gamma}} {f_{max} - f_{min}}, \ \forall j \in [1,m]
\label{eq:ei_def}
\end{equation}

The statistical distribution of $E_i$ is presented with a box plot formalism. A
box plot groups the data through different quantiles: the bottom and the top of
the box represent respectively the value of the first and third quartiles,
whereas the horizontal line inside the box is the median (second quartile) and
the diamond the mean. The vertical lines indicate the data between the 5th
percentile and the first quartile and betwwen the third quartiles and the 95th
percentile. Finally the outliers are plotted as dots.

\subsection{Two-Dimensional RAE2822 Transonic Airfoil}
The viscous and turbulent flow around a RAE2822 airfoil has been widely studied
in the literature both numerically and experimentally~\cite{Cook1977,
Haase1993euroval}. The feature of prime interest of this test case is that the inflow conditions govern
the flow regime, leading to the appearance of shock waves. The detection and the
clear separation of these regimes represent the main challenge for the model,
demonstrating the capability of the LDM to deal with real and complex physics
compared to the classical method.
\subsubsection{Computational configuration}
\begin{figure} 
\centering
  \begin{subfigure}{.5\textwidth}
	\centering  
	\includegraphics[width=0.9\linewidth]{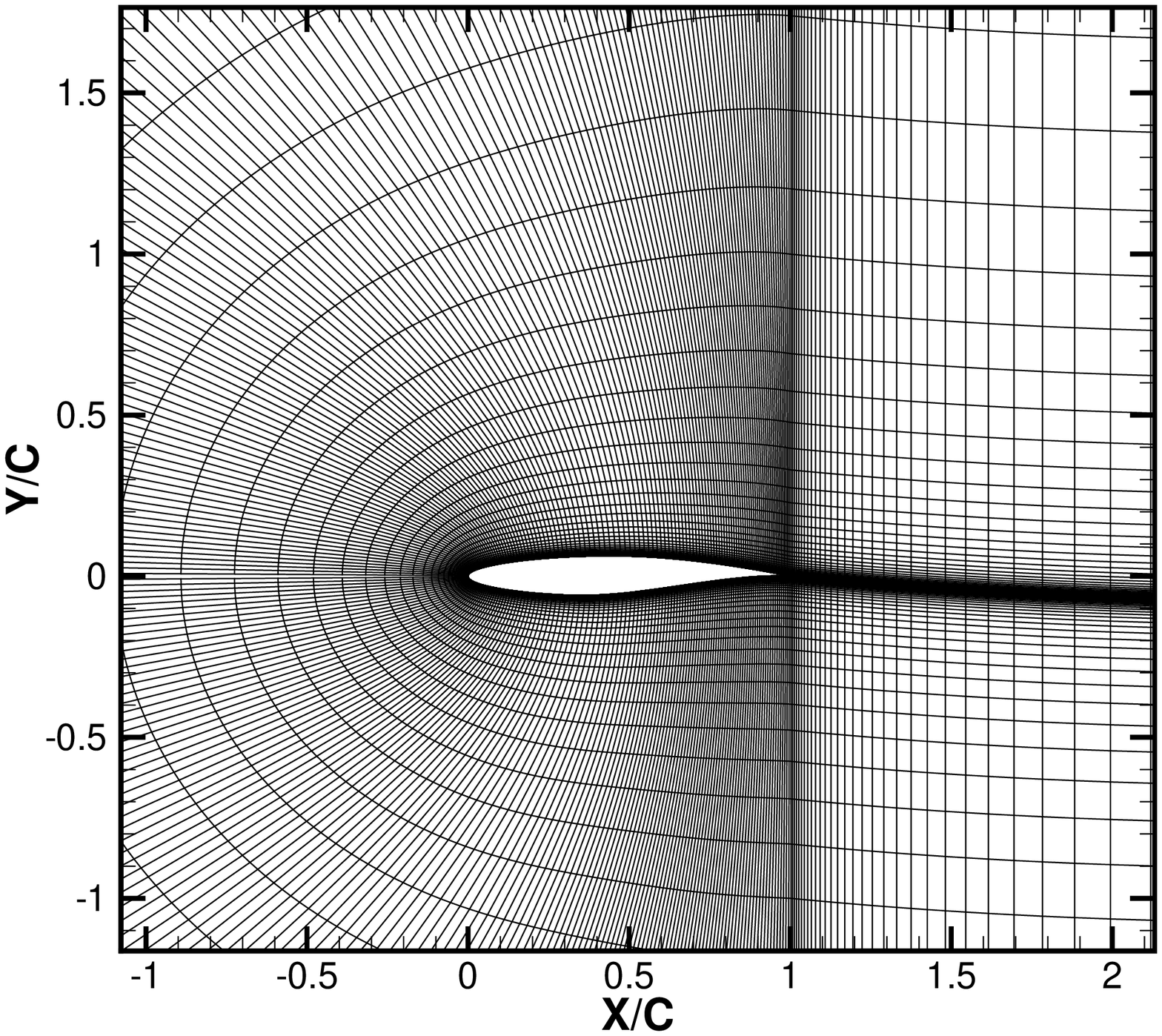}
	\caption{View of the computational grid.}
	\label{fig:rae_grid}  
  \end{subfigure}%
  \begin{subfigure}{.5\textwidth}
	\centering  
	\includegraphics[width=0.9\linewidth]{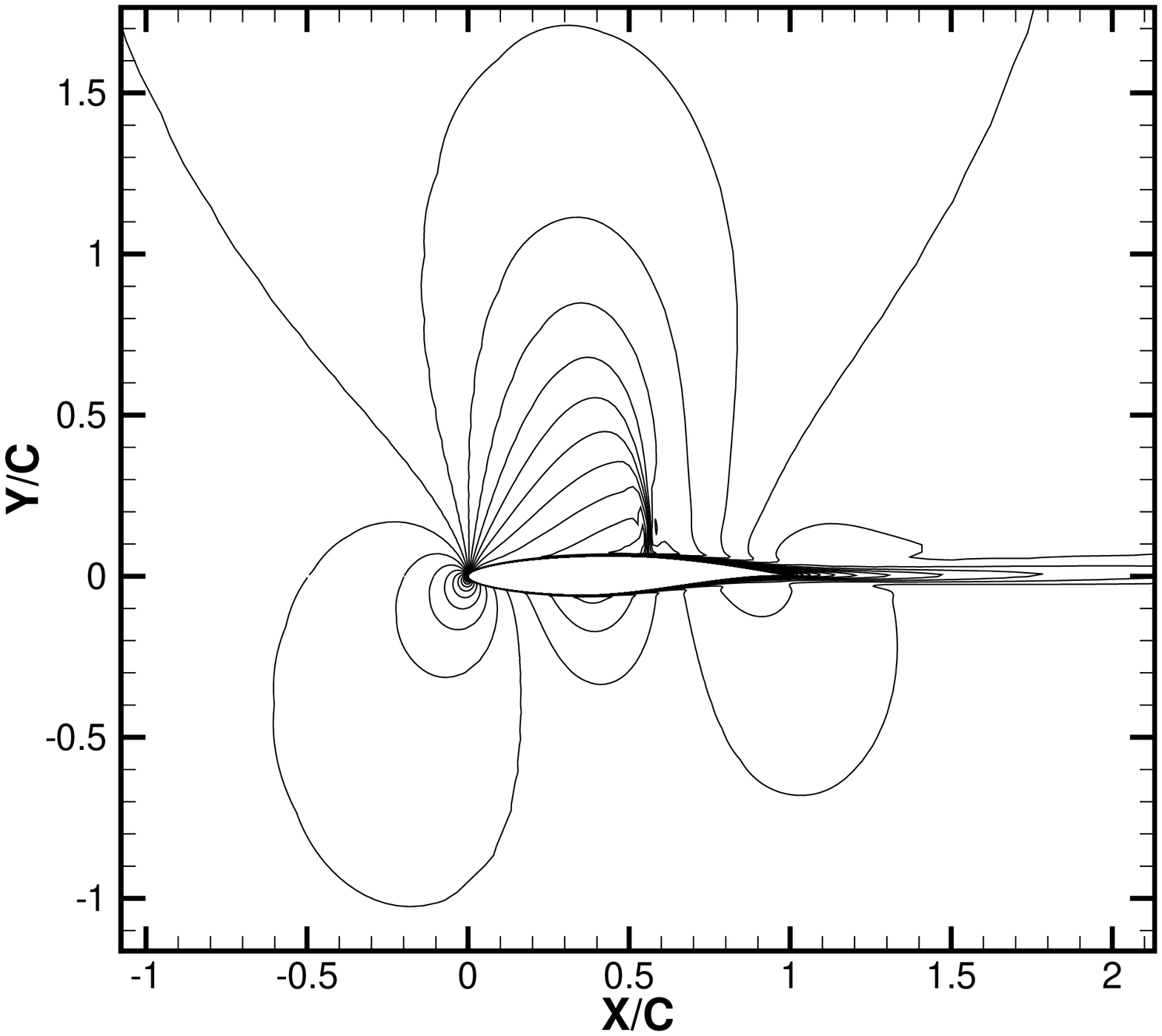} 
	\caption{Mach number contours (M=0.734 and $\alpha$=2.79).}
	\label{fig:rae_mach_contours}
  \end{subfigure}
\caption{Flow around the RAE2822 airfoil.}
\label{fig:rae2822} 
\end{figure}

The high-fidelity computations are carried out using the cell-centered
finite-volume solver elsA~\cite{Cambier2008}. It has been developed at ONERA and
solves the compressible Reynolds Average Navier Stokes (RANS) equations on
structured grids. From the numerical point of view, the classical second order
central scheme of Jameson, Schmidt and Turkel~\cite{Jameson1981} is used for the
space discretization. The time integration is performed with the backward Euler
implicit scheme: the algebraic system is linearized with the LU-SSOR implicit
method~\cite{Yoon1988}. The turbulence modeling is ensured by the model of
Spalart and Allmaras. A $2D$ mesh containing $23,010$ points is used, as
illustrated in the~\figurename{\ref{fig:rae_grid}}. This test case has been
successfully validated on a well-known regime flow~\cite{Cook1977}
(\figurename{\ref{fig:rae_mach_contours}}). The chord of the airfoil is written
$C$, $X$ refers to the horizontal coordinate and $Y$ refers to the vertical
coordinate. 
\subsubsection{Input parameter space, quantities of interest and settings of
the surrogate model}
Three freestream parameters are considered as varying for
this application: the flight speed, the angle of attack $\alpha$, and the altitude $h$. Their
variations are resumed in Table~\ref{tab:free_stream}. These dimensional
parameters are nondimensionalized and are expressed respectively as the Mach
Number $M$, angle of attack $\alpha$ (no change) and Reynolds number. One can
note that the latter is impacted by the altitude variations but ensured to be
maintained in a given interval in order to have a sufficiently resolved boundary layer.
\begin{table}
\centering
\begin{tabular}{cc}
\hline\hline
Freestream variable& Amplitude of variation  \\\hline
Flight speed (m.s$^{-1}$)& 88.5 - 269\\
Angle of attack ($^\circ$)&  0.5 - 3.0 \\
Altitude (m)& 1000 - 11,000\\
\hline\hline
\end{tabular}
\caption{\label{tab:free_stream} Freestream conditions.}
\end{table}
The atmosphere is modeled by the International Standard Atmosphere of the
International Civil Aviation Organization~\cite{ISO1978}. It assumes that the
air is a perfect gas and that the atmosphere can be divided into layers with a
linear distribution of temperature against the altitude. The temperature $T$
and the density $\rho$ can be directly expressed in function of the altitude:
\begin{align}
T &= T_0 - L \, h\\
\rho &= \frac{p_0 (1- \frac{L \, h}{T_0})^{\frac{g}{rL}}}{r(T_0-L \, h)}
\end{align} 
with $p_0$ and $T_0$ the pressure and temperature at sea level, $L$ the
temperature lapse rate, $r$ the specific gas constant of air, and $g$ the
acceleration due to the gravity. 
The quantities of interest of the simulation are the pressure
coefficient $C_p$ and the friction coefficient $C_f$ defined by.
\begin{align}
C_p = \frac{p-p_{\infty}}{\frac{1}{2}\rho_{\infty}U_{\infty}^2}\\
C_f = \frac{\tau_w}{\frac{1}{2}\rho_{\infty}U_{\infty}^2}
\end{align}
where $p$ is the static pressure, $\tau_w$ the wall shear stress, $p_{\infty}$,
$\rho_{\infty}$ and $U_{\infty}$ respectively the static pressure, the density
and the velocity in the freestream.

As regards the DOE, 30 samples of an Halton sequence form the training set,
following the rule of thumb $10d$~\cite{Loeppky2009}. The sampling of the LDM is
divided into two parts. An initial DOE mixing subsonic and transonic snapshots
explores uniformly the parameter space with an Halton sequence of 15 samples.
The last 15 samples follow the resampling process described in the previous
section. The flow is assumed to be driven by two different flow regimes.
Therefore, the number of clusters is set to two for the clustering step. A test
set has been built from 300 snapshots of a Sobol sequence in order to assess the
LDM.
\begin{figure}
\centering
  \begin{subfigure}{.49\textwidth}
	\centering  
	\includegraphics[width=1.05\linewidth]{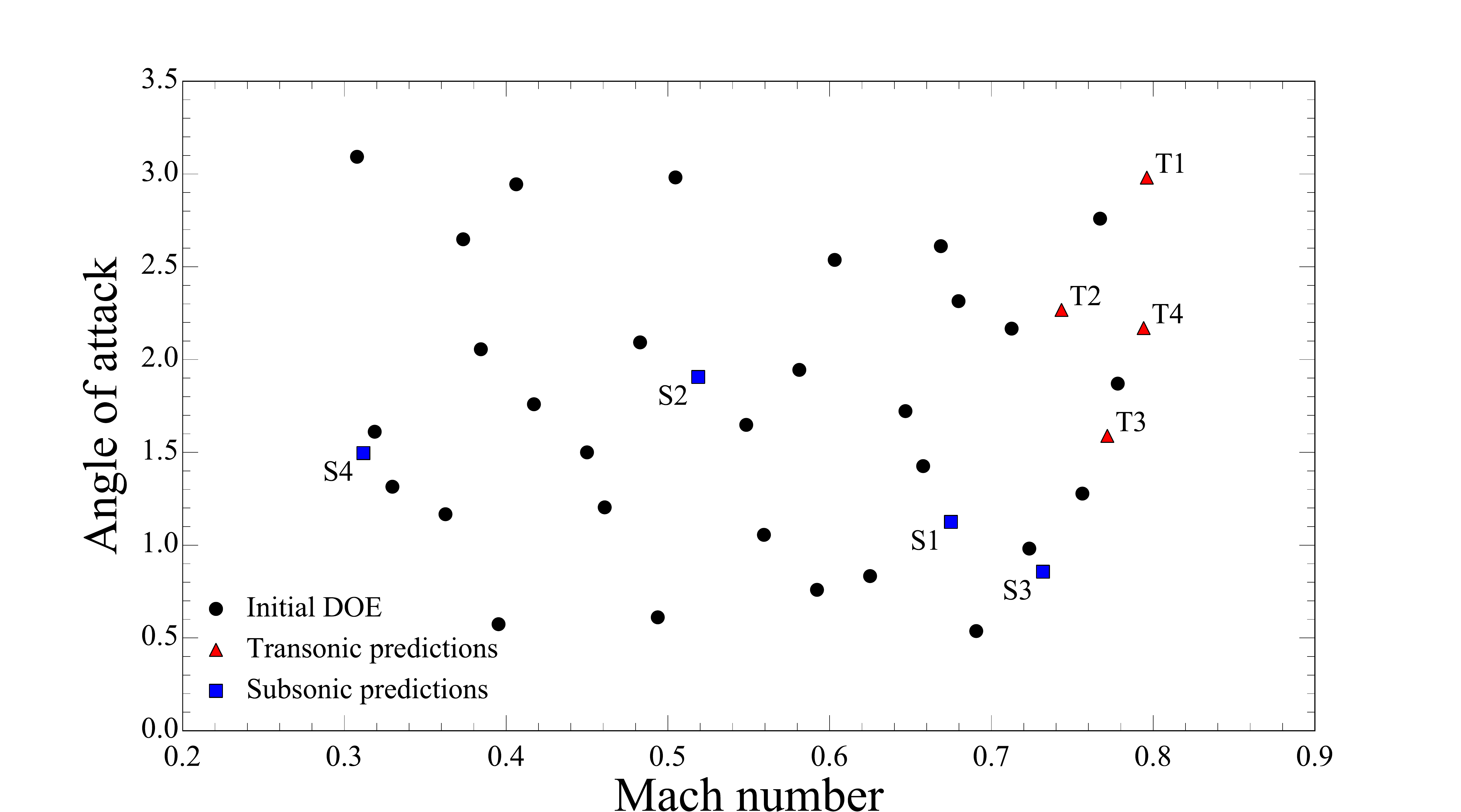}
	\label{fig:halton}  
  \end{subfigure}%
  \begin{subfigure}{.49\textwidth}
	\centering  
	\includegraphics[width=1.05\linewidth]{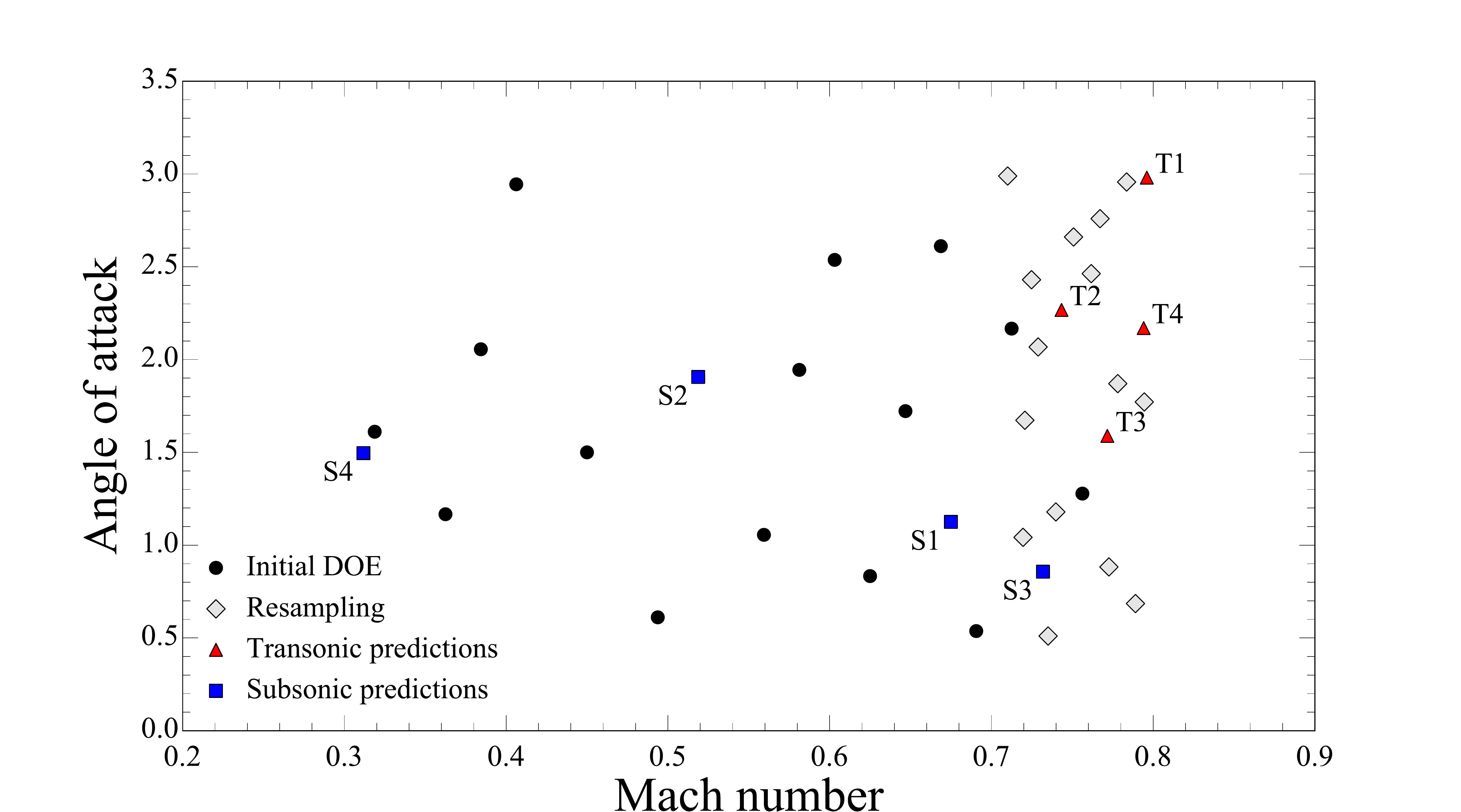}
	\label{fig:cp_profile_cruise}
  \end{subfigure}
\caption{Thirty samples from a classical Halton sequence (left) and resampling
technique (right). The eight illustrative predictions are also identified (with
blue squares and red triangles).}
\label{fig:resampling}
\end{figure}

\subsubsection{Analysis of the surrogate model building process}
\begin{figure}
\centering
  \begin{subfigure}{.5\textwidth}
	\centering  
	\includegraphics[width=1.1\linewidth]{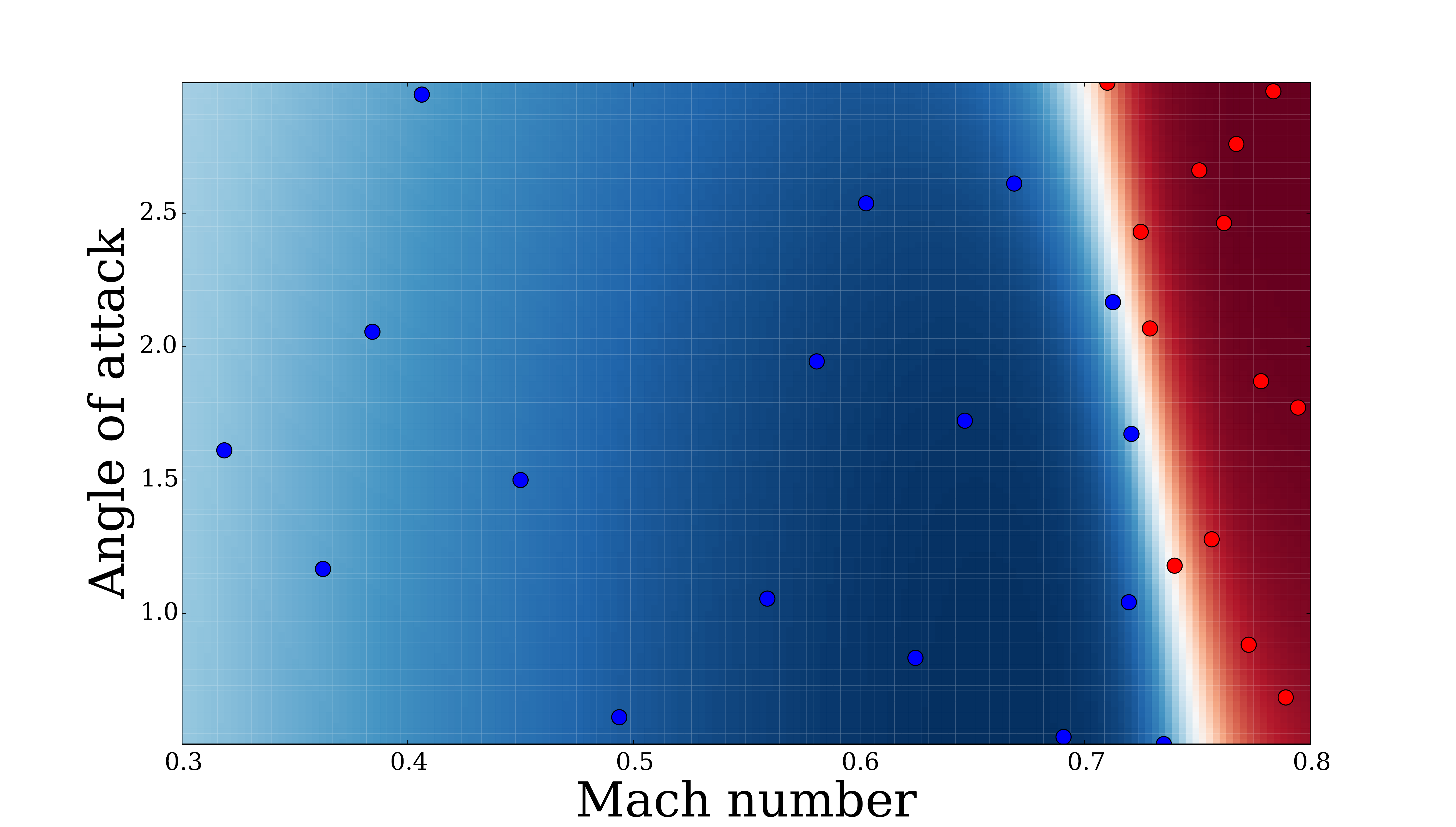}
	\caption{Training set}
  \end{subfigure}%
  \begin{subfigure}{.5\textwidth}
	\centering  
	\includegraphics[width=1.1\linewidth]{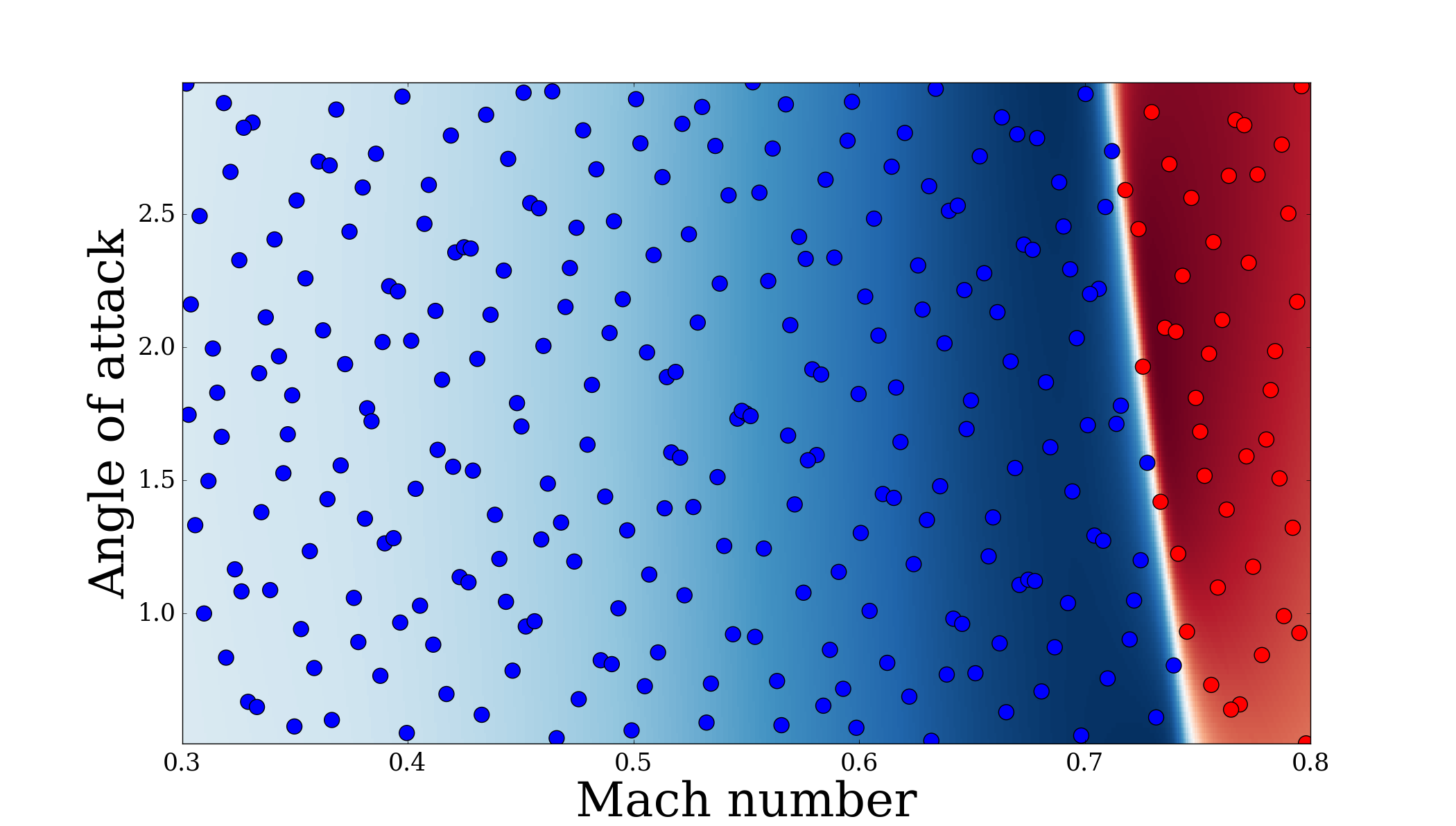}
	\caption{Test set}
	\label{fig:test_input_space_decomposition}
  \end{subfigure}
\caption{Input space decomposition. Each color corresponds to a cluster, the
blue one is the subsonic regime and the red one is the transonic regime.}
\label{fig:input_space_decomposition}
\end{figure}

The two training sets exhibit interesting differences. The Halton
approach explores uniformly the full input space, whereas the resampling process
focuses on a low-dimensional High Mach number region, as depicted
in~\figurename{\ref{fig:resampling}} for the $C_p$. Indeed, the clustering phase
automatically identifies the subsonic and the transonic snapshots thanks to the
shock sensor. The supervised algorithm decomposes the input space parameter
allowing to determine the separation of the two physical regimes in the input
parameter space, as illustrated in
the~\figurename{\ref{fig:input_space_decomposition}} with both the training and
testing sets. These two clusters can be interpreted as the subsonic and the
transonic regions. It can be observed that the boundary is mainly influenced by
the Mach number but also slightly by the angle of attack. Thus, the resampling
process has increased the density of samples in the transonic regime, improving
the accuracy of the model where the predictions are more challenging. One can
note that the boundary region is well defined by a thin region of probability
between 40\% and 60\%.
Thus, the interface model defined by the global model is only applied in this
small region.

The model can also be analyzed from a dimension reduction point of view with the
Table~\ref{tab:rae_compressibility}. For a given POD energy ratio of 0.99\%, the
classical method reduces the dimension of the $30$ snapshots with 10 modes and
an entropy of 0.37, whereas the LDM identifies clearly a POD basis with a low
entropy and a POD basis with a large one. The highest nonlinear cluster shows
an entropy values of 0.63, 70\% bigger than the linear cluster and the classical
method.
\begin{table}
\centering
\begin{tabular}{cccc}
\hline\hline
Method& Number of snapshots& Number of modes& Entropy  \\\hline
Classical& 30& 7& 0.37\\
LDM (transonic regime)&  13& 9& 0.63\\
LDM (subsonic regime)& 17& 6& 0.36\\
\hline\hline
\end{tabular}
\caption{\label{tab:rae_compressibility} Required modes of the different
methods for the RAE2822.}
\end{table}
It means the data of the subsonic POD basis and of the classical method can be
highly compressed. Conversely, the transonic POD basis requires 9 modes
for 13 snapshots. Thus, the active resampling provides extra non-redundant
information to the LDM compared with the classical method.

\begin{figure}
\centering
\begin{subfigure}[h]{1\textwidth}
\centering 
\includegraphics[width=1.0\textwidth]{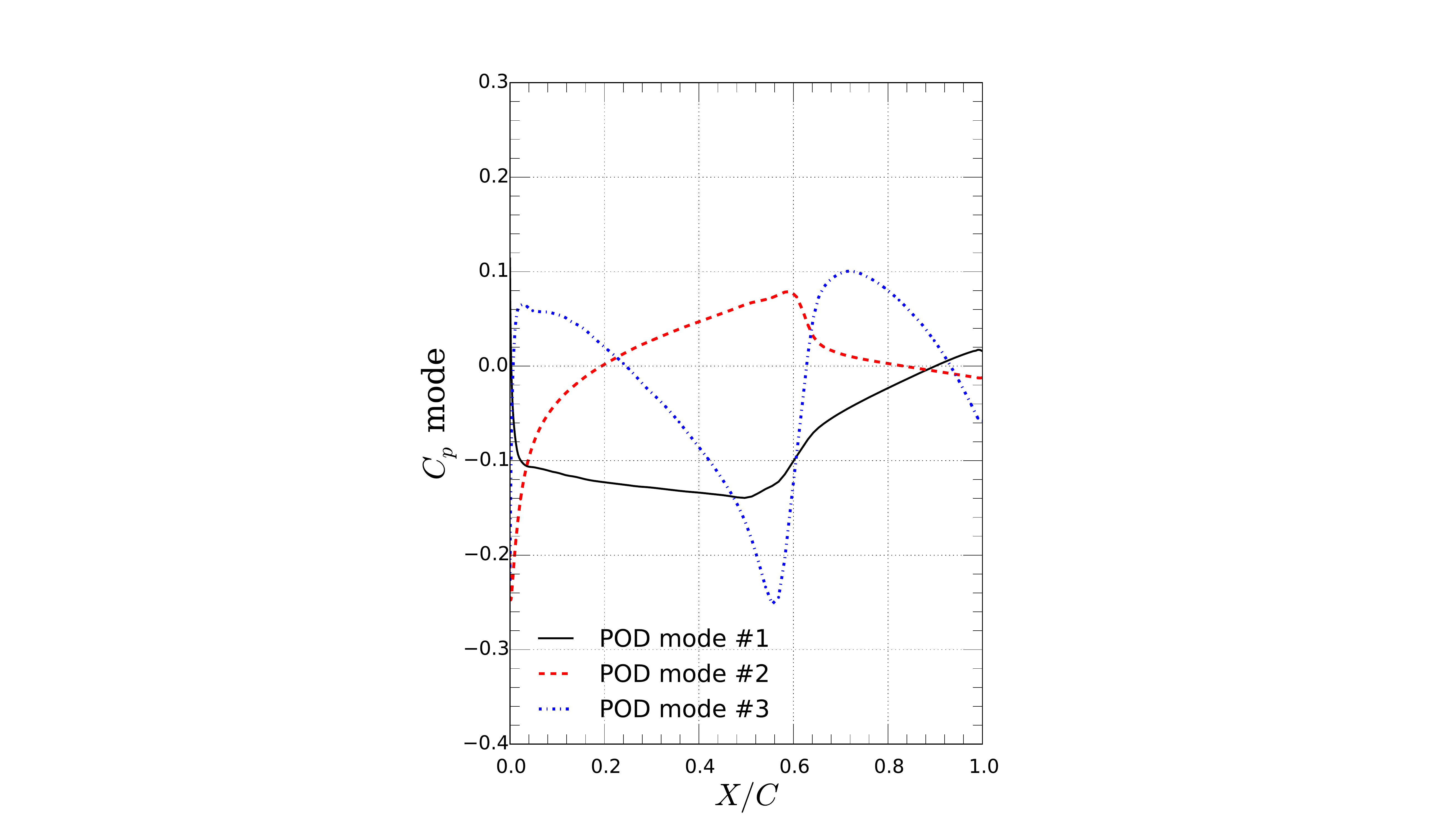} 
\caption{POD modes for the classical method.}
\label{fig:pod_modes_classical}
\end{subfigure}
\begin{subfigure}[h]{1\textwidth}
\centering
\includegraphics[width=0.95\textwidth]{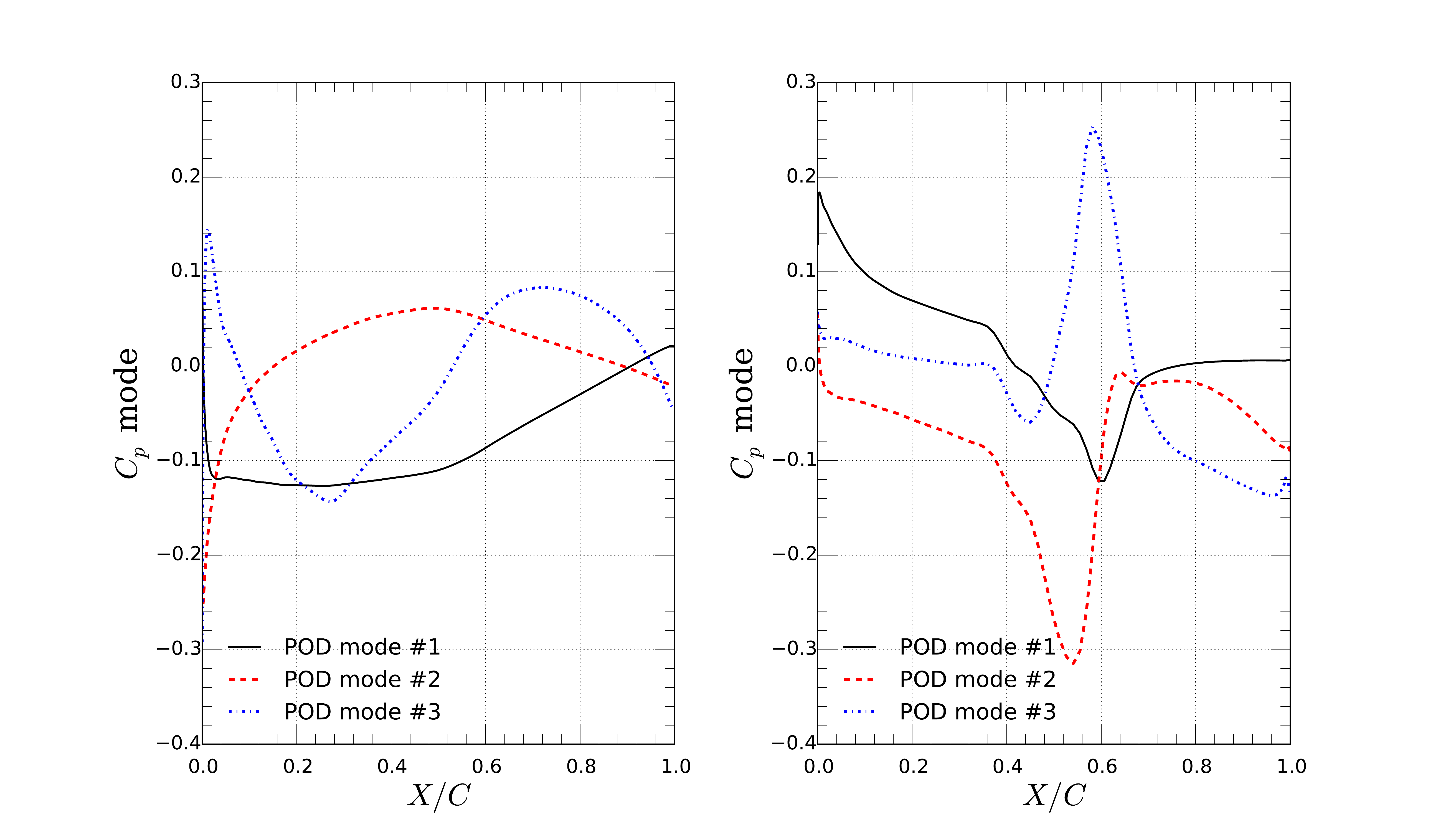}
\caption{POD modes of the $C_p$ for the LDM, subsonic (left) and transonic
(right).}  
\label{fig:pod_modes_ldm}
\end{subfigure}
\caption{POD modes of the RAE2822 simulations.}
\label{fig:rae_pod_modes}
\end{figure}

The shape of the modes can also help to understand the behavior of the different
models. As regards the classical method, only discontinuous features emerge
clearly from the global POD as illustrated
in~\figurename{\ref{fig:pod_modes_classical}}, where POD modes shown only
dominant discontinuities and no-moving shocks. Therefore all the reduced
coordinates associated with subsonic snapshots must exactly cancel out the
discontinuities of the modes. For this reason, the prediction of the reduced
coordinates for snapshots in the subsonic region can be sensitive to
interpolation errors leading to the appearance of ``residual'' shocks. Thus, the
clear separation of the regimes in the POD domain represents a major asset
inherent in the LDM. Figure~\ref{fig:pod_modes_ldm} shows the first three modes
of the two POD bases. High nonlinearities arises for the transonic regime
whereas the subsonic region highlights similarities with the modes of the classical
method. The only difference is that the subsonic modes are smoother.  

\begin{figure}
\centering
\begin{subfigure}[h]{1\textwidth}
\centering
\includegraphics[width=\textwidth]{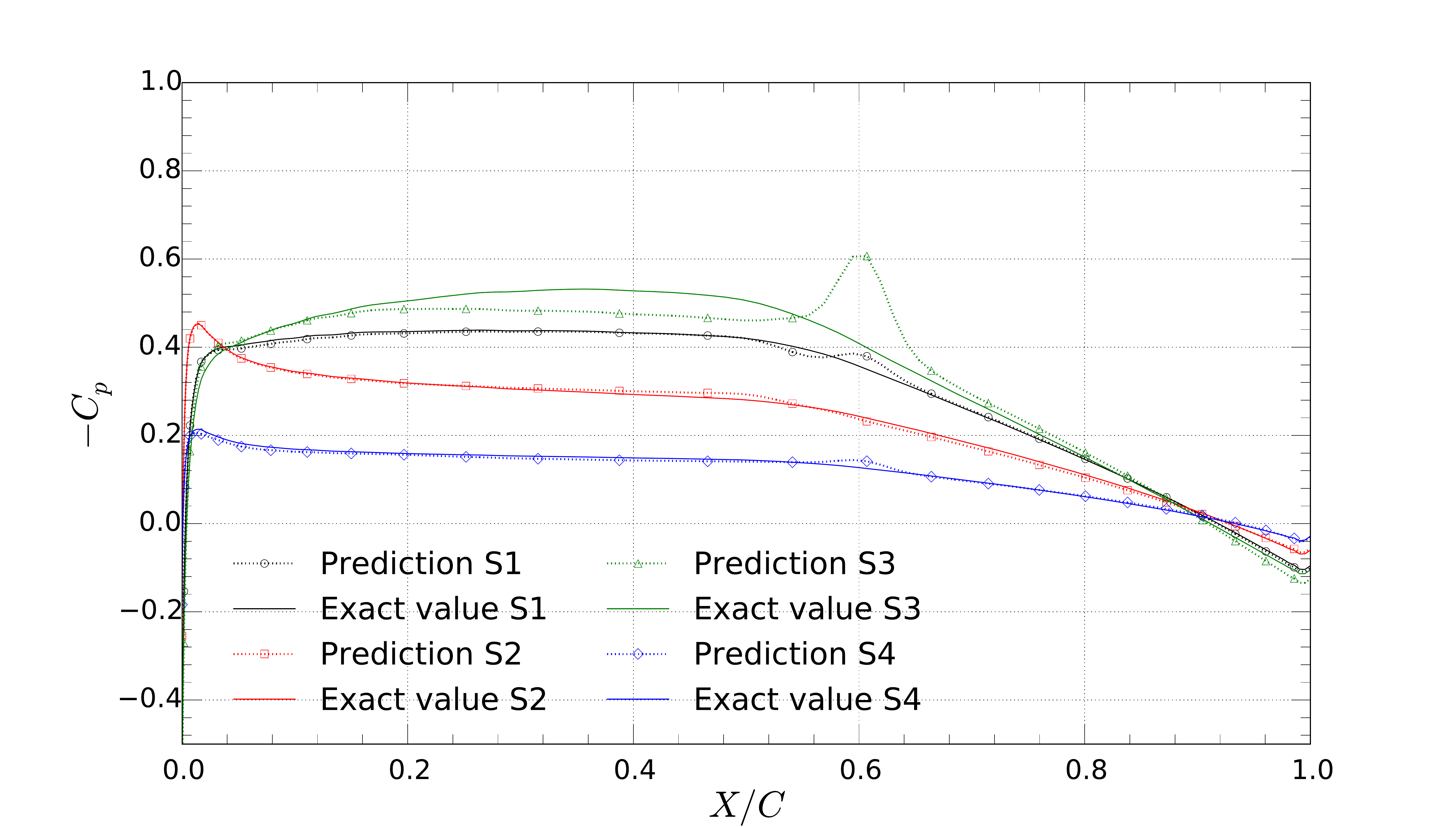} 
\caption{Classical model}
\label{fig:cp_profile_subsonic_classical}
\end{subfigure}

\begin{subfigure}[h]{1\textwidth}
\centering
\includegraphics[width=\textwidth]{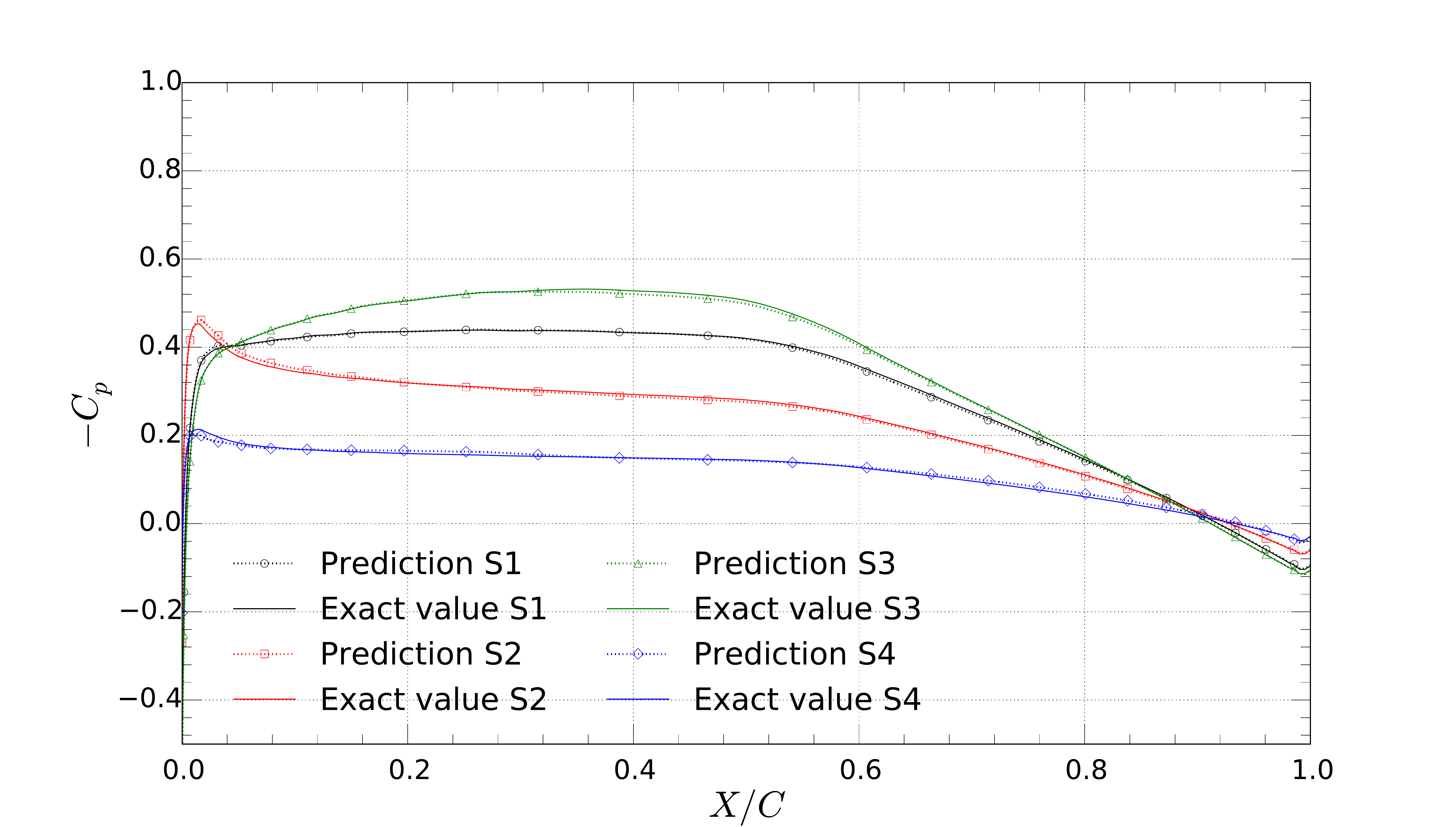}
\caption{LDM model}
\label{fig:cp_profile_transonic_classical}
\end{subfigure}

\caption{$C_p$ profiles in the subsonic regime.}
\label{fig:cp_profile_subsonic}
\end{figure}

\begin{figure}
\centering
\begin{subfigure}[h]{1\textwidth}
\centering
\includegraphics[width=\textwidth]{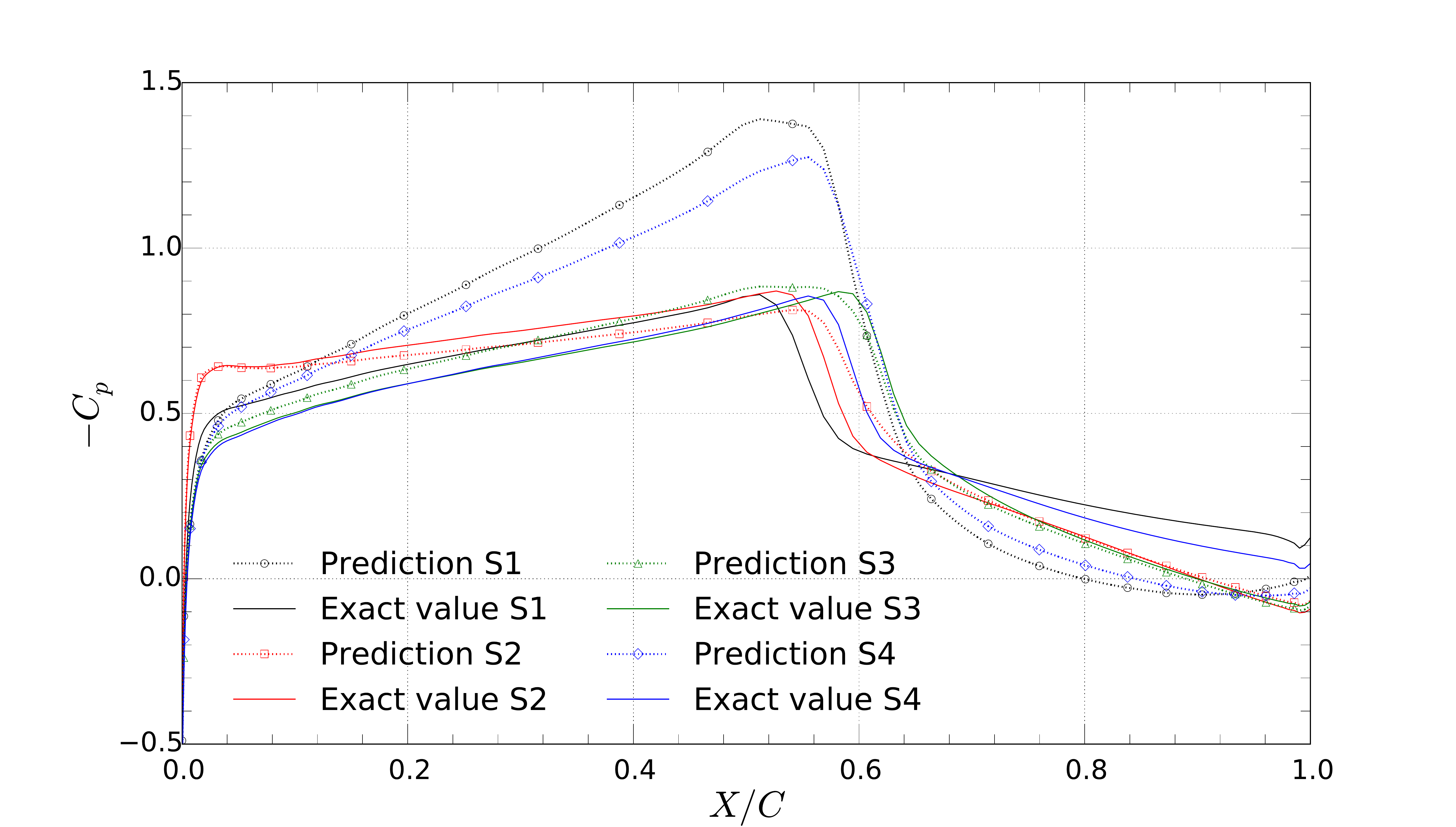} 
\caption{Classical model}
\label{fig:cp_profile_transonic_classical}
\end{subfigure}

\begin{subfigure}[h]{1\textwidth}
\centering
\includegraphics[width=\textwidth]{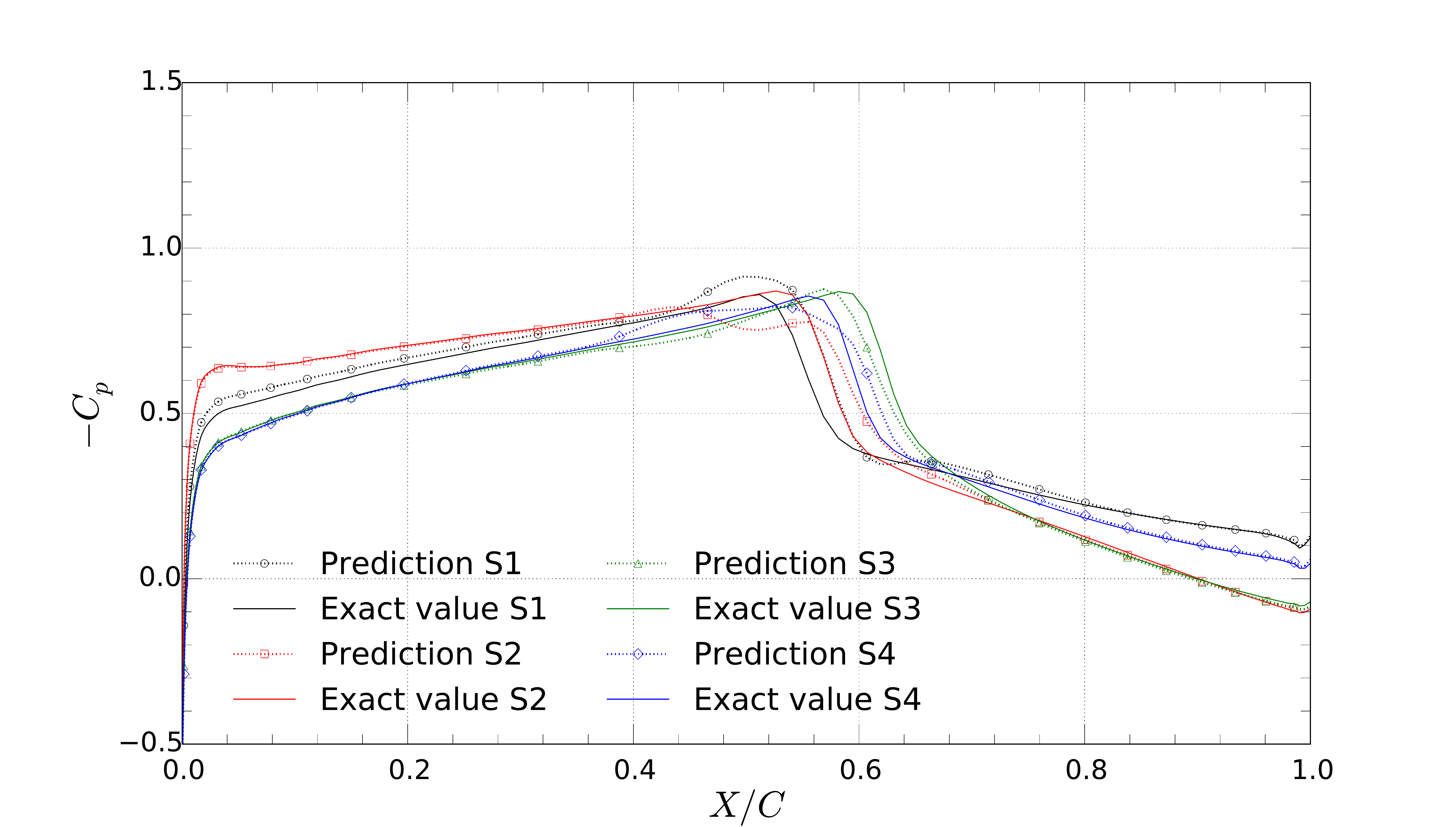}
\caption{LDM model}
\label{fig:cp_profile_transonic_ldm}
\end{subfigure}

\caption{$C_p$ profiles in the transonic regime.}
\label{fig:cp_profile_transonic}
\end{figure}

These statements are confirmed by looking at $C_p$ profiles, comparing classical
and LDM methods. Eight predictions have been computed for an illustrative
purpose and are grouped by physical regime (S for subsonic and T for transonic).
Their coordinates in the parameter space are summarized in the
Table~\ref{tab:prediction_points} and their repartition is illustrated
in~\figurename{\ref{fig:resampling}}.
One can observe that for the subsonic regime
in~\figurename{\ref{fig:cp_profile_subsonic}}, the classical method induces
residual shocks, certainly due to errors of prediction on the reduced
coordinates coupled with sharp POD mode not adapted to subsonic flows. On the
contrary, the residual shocks are filtered by the LDM as lower discontinuities
are present in the training snapshots building the POD basis. It leads to final
predictions less sensitive to errors on the reduced coordinates. As regards
$C_p$ profiles shown in the transonic regimes
in~\figurename{\ref{fig:cp_profile_transonic}}, the LDM shows improved accuracy.
In particular, the modeling of the shock waves gives an insight into the
behavior of the two models. Significant discrepancies in term of shock
displacement and shock amplitude occur for the classical POD/GPR method.
Indeed, the combination of Angle of attack and Mach number influences directly
the localization of the steady shock wave and its amplitude. However this
behavior can be accurately caught only if the training set contains a sufficient
amount of snapshots with shocks, which is not the case for the classical POD/GPR
methods. The same trend is observed for the $C_f$ profiles
in~\figurename{\ref{fig:cp_profile_transonic}}.
\begin{table}
\centering
\begin{tabular}{cccc}
\hline\hline
Predictions& Mach Number & Angle of attack ($^\circ$) & Altitude (m)  \\\hline
S1& 0.675&   1.125&   8500\\
S2&   0.519&   1.906&   1625\\
S3&  0.722&   1.047&   1937\\
S4&   0.312&   1.496&   5453\\
T1&  0.675&  1.125&   8500\\
T2&  0.519&   1.906&   1625\\
T3& 0.722&   1.047&   1937\\
T4&  0.312&   1.496&   5453\\
\hline\hline
\end{tabular}
\caption{\label{tab:prediction_points} Coordinates in the parameter space of the
eight illustratives predictions.}
\end{table}
 
\begin{figure}
\centering
\begin{subfigure}[h]{1\textwidth}
\centering
\includegraphics[width=\textwidth]{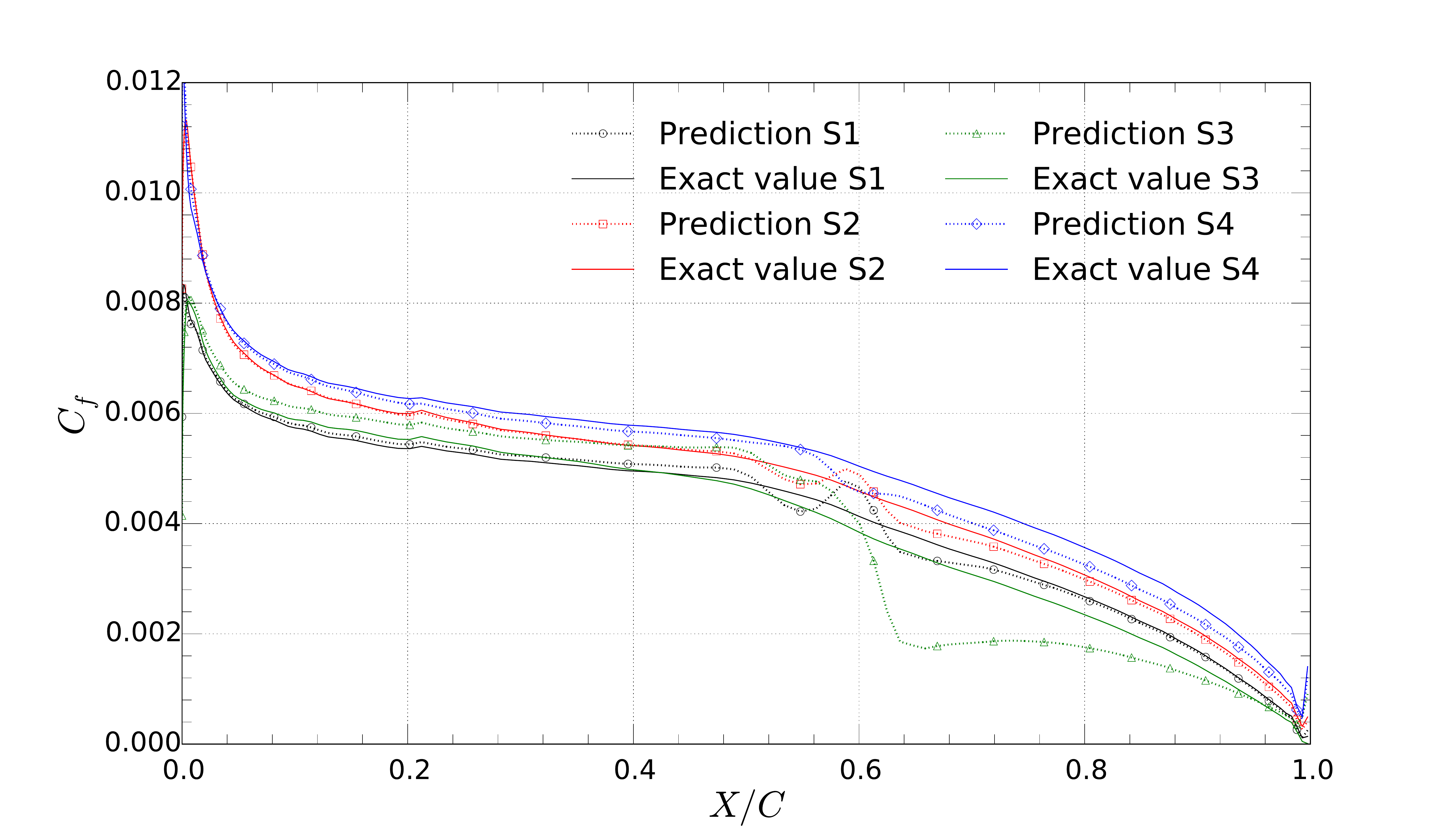} 
\caption{Classical model}
\label{fig:cf_profile_subsonic_classical}
\end{subfigure}

\begin{subfigure}[h]{1\textwidth}
\centering
\includegraphics[width=\textwidth]{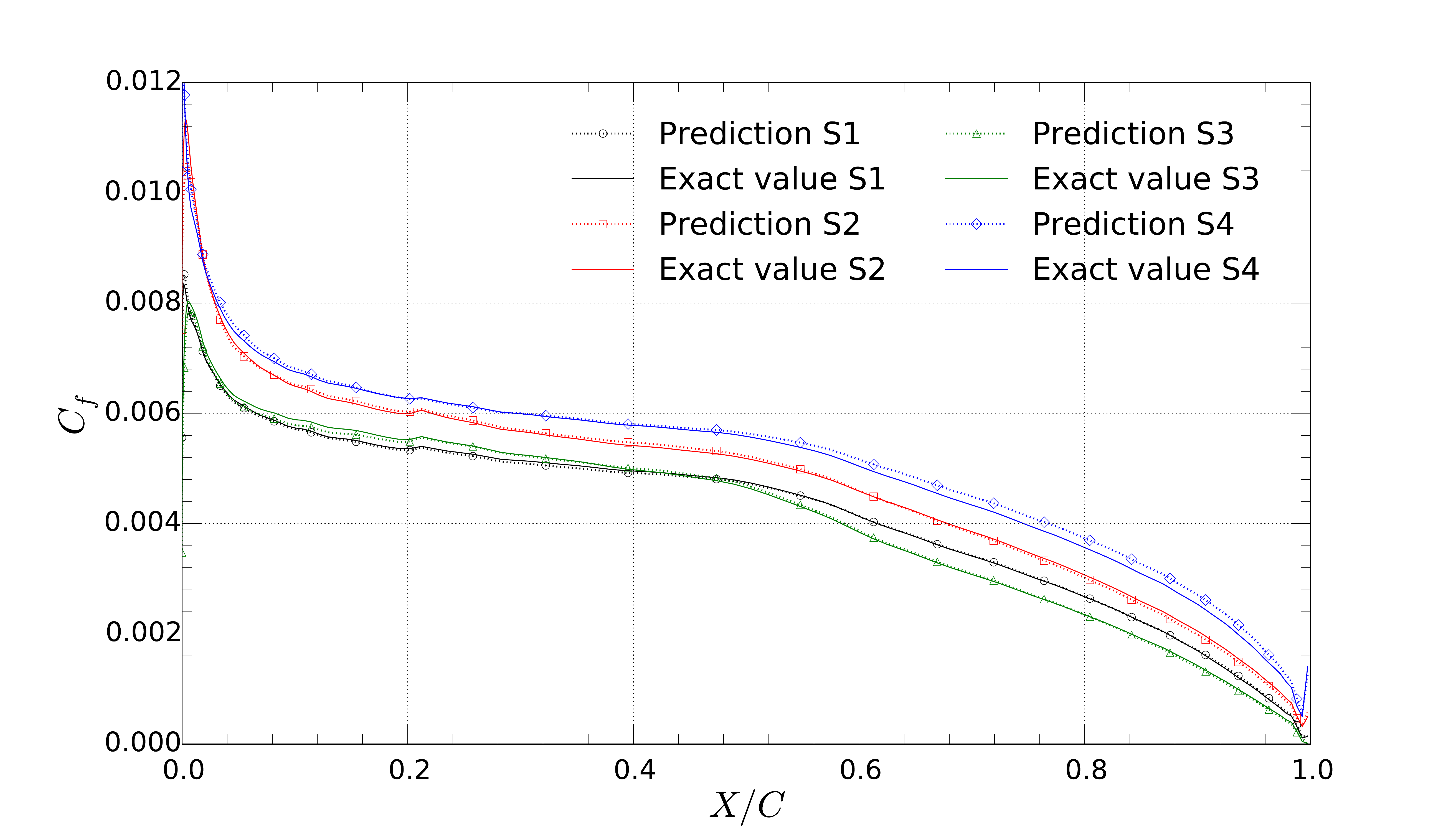}
\caption{LDM model}
\label{fig:cf_profile_transonic_classical}
\end{subfigure}

\caption{$C_f$ profiles in the subsonic regime.}
\label{fig:cf_profile_subsonic}
\end{figure}

\begin{figure}
\centering
\begin{subfigure}[h]{1\textwidth}
\centering
\includegraphics[width=\textwidth]{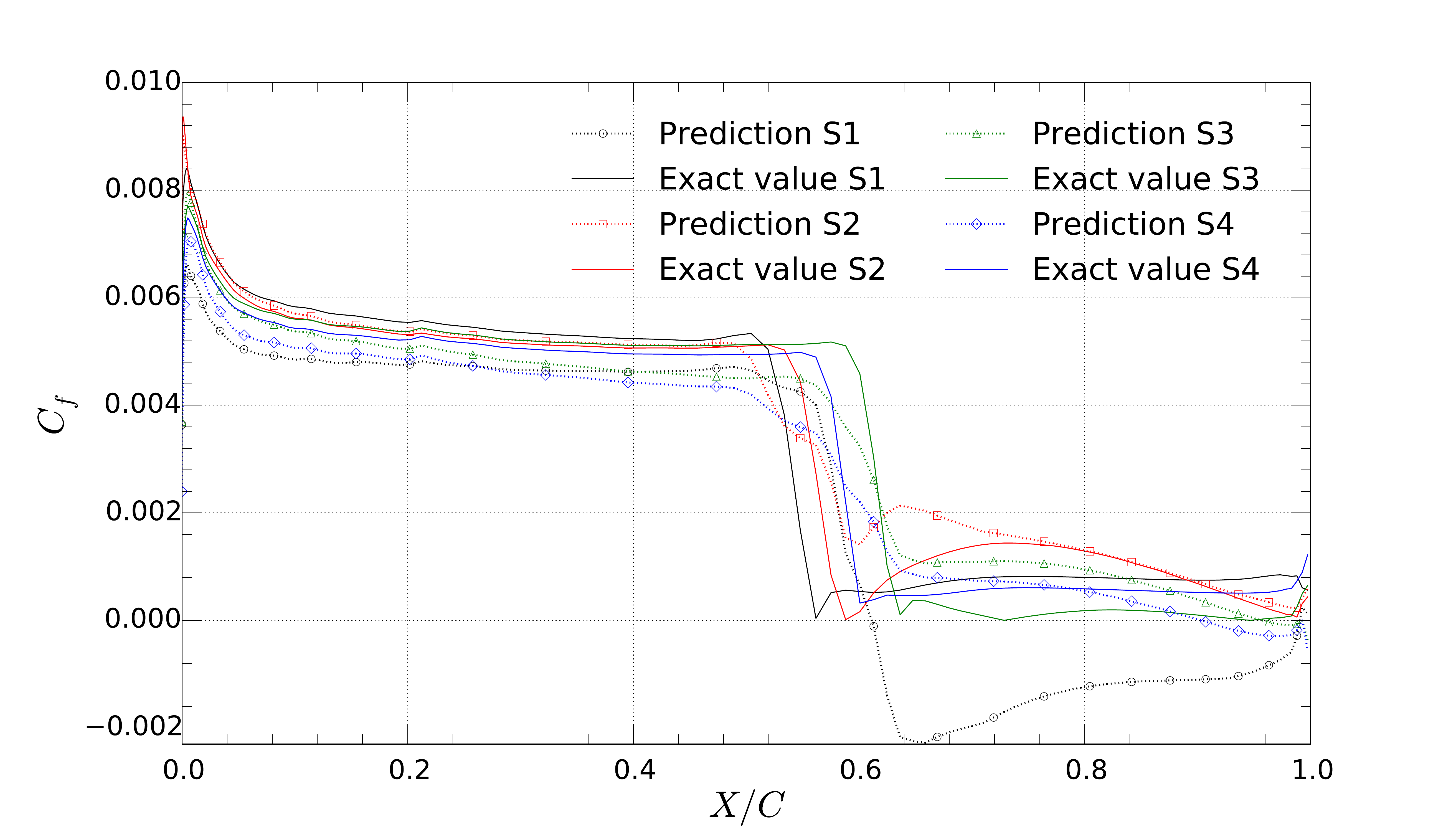} 
\caption{Classical model}
\label{fig:cf_profile_transonic_classical}
\end{subfigure}

\begin{subfigure}[h]{1\textwidth}
\centering
\includegraphics[width=\textwidth]{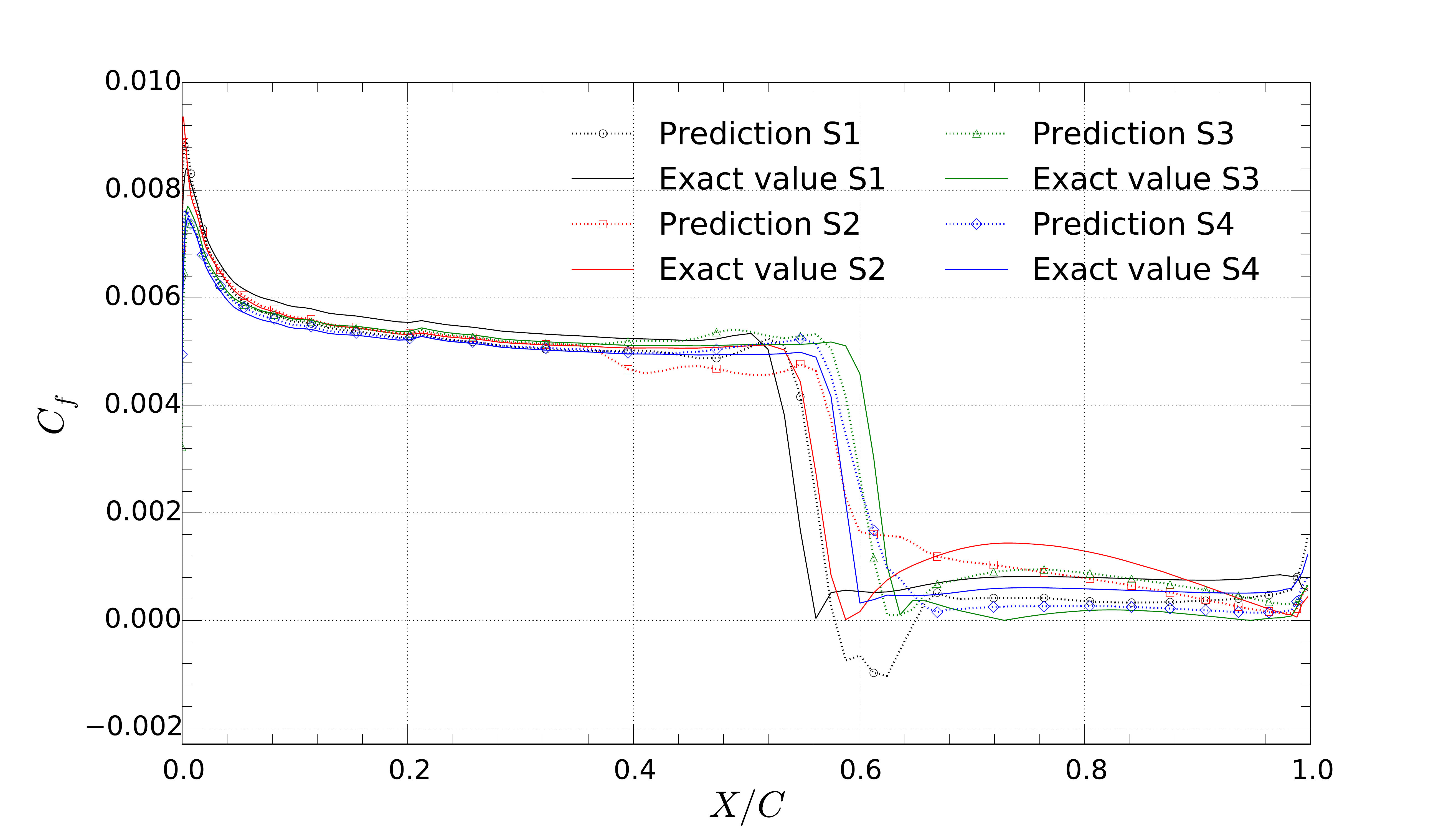}
\caption{LDM model}
\label{fig:cf_profile_transonic_ldm}
\end{subfigure}

\caption{$C_f$ profiles in the transonic regime.}
\label{fig:cf_profile_transonic}
\end{figure}

\subsubsection{Accuracy of the model}
\begin{figure}
\centering
\begin{subfigure}[h]{1\textwidth}
\centering
\includegraphics[width=\textwidth]{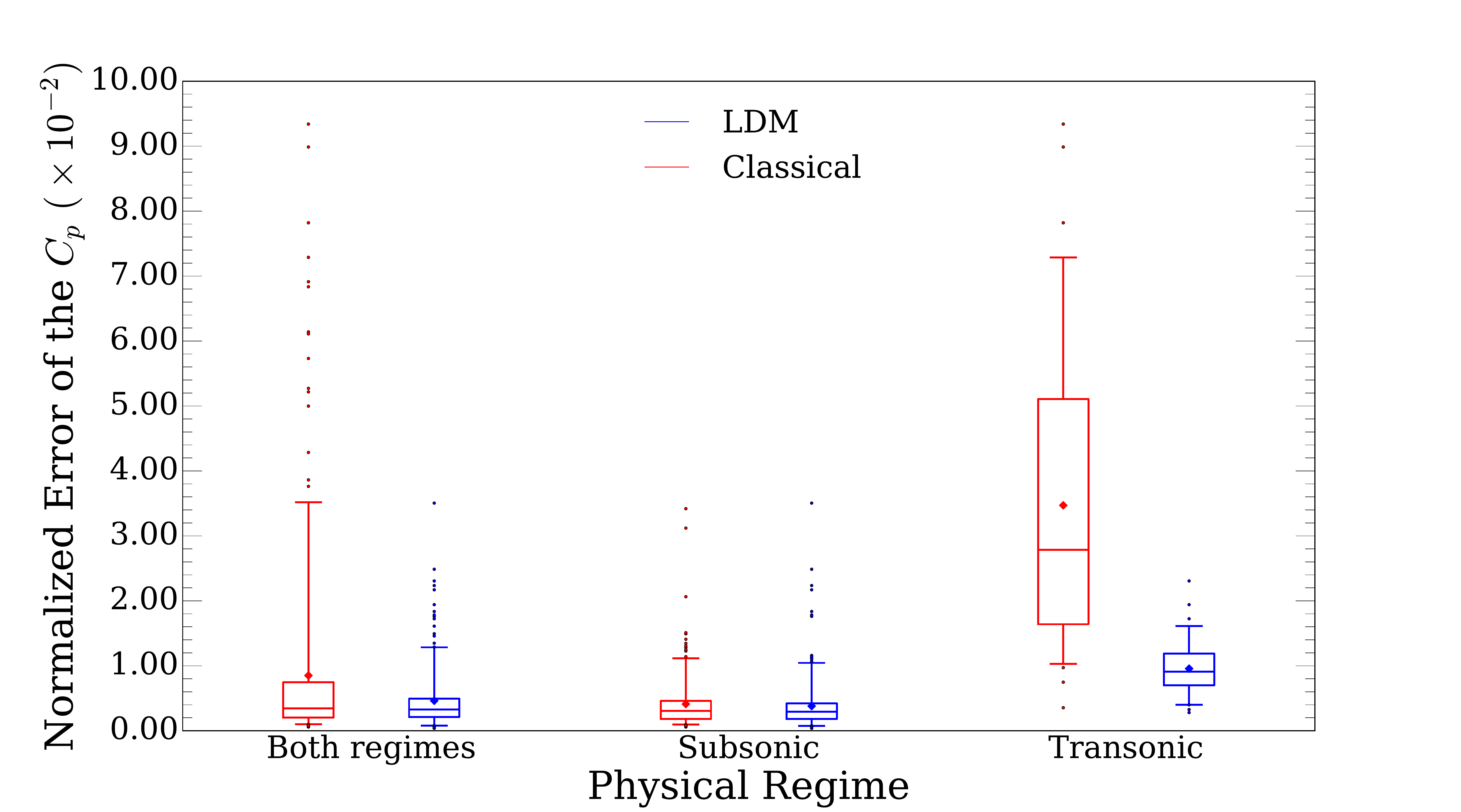} 
\label{fig:cp_box_plot}
\end{subfigure}

\begin{subfigure}[h]{1\textwidth}
\centering
\includegraphics[width=\textwidth]{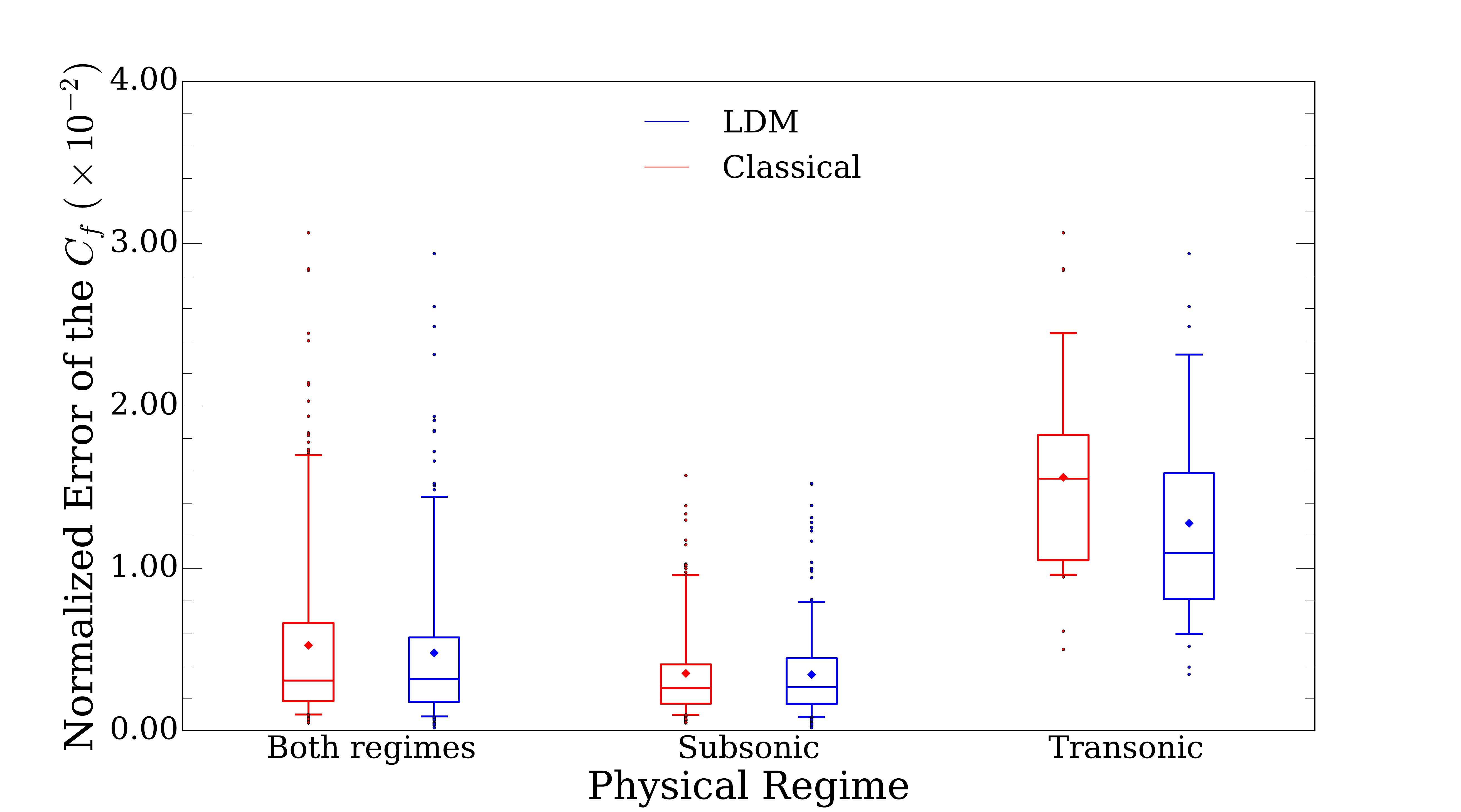}
\label{fig:cf_box_plot}
\end{subfigure}
\caption{Accuracy of the surrogate model in terms of $C_p$ and $C_f$.}
\label{fig:surrogate_boxplot}
\end{figure}

In this section, a more detailed look is given to the analysis of the model
accuracy. Figure~\ref{fig:surrogate_boxplot} displays the comparison of the normalized
error in term of $C_p$ and $C_f$ for both methods between the predictions and
the test set. The results are presented with a box plot formalism and three
different phases has been considered:
\begin{itemize}
  \item The full domains contains all the samples of the testing set.
  \item The subsonic regime is only composed of the testing samples identified
  as subsonic, in blue dots in~\figurename{\ref{fig:test_input_space_decomposition}}.
  \item The transonic regime encompasses the other snapshots, shown as red dots
  in~\figurename{\ref{fig:test_input_space_decomposition}}.
\end{itemize}
A significant improvement in the accuracy is induced by the LDM for the $C_p$.
As regards the full domain, the normalized error decreases dramatically for all
the statistical characteristics of the box plot. In particular, the extreme
value of the LDM reaches the same level as the 95\% error of the classical
method, illustrating a large reduction of the model variability. The box plots
for subsonic and transonic regimes provide a closer look at the repartition of
the error. It clearly appears that the LDM improves the predictions at transonic
regime, explained by the increase of the sample density. On the other
hand, the subsonic regime is very slightly impacted compared with the classical
method, although samples has been removed.

The Table~\ref{table:rae_surrogate_results} provides a more global view of the
error with the spatial average of the $Q_2$ and NRMSE. Whatever the quantity
measuring the error, the same trend is observed for the two quantities of
interest: the LDM significantly improves the accuracy of the predictions
compared with the classical method. There are, however, legitimate doubts as to
the value of the $\langle Q_2 \rangle_\Gamma$ for the $C_f$ which is below zero
for the transonic regime. Several explications can be given. First of all, the
$\langle Q_2 \rangle_\Gamma$ is spatially averaged and the value of $C_f$ can be
very close to zero after the shock. Thus, some values of $Q_2$ falls far below zero in this
region, impacting directly the average of the predictivity coefficient. Then,
the skin friction coefficient is also more challenging to predict due to its
higher dependance to the altitude and higher nonlinearity.




\begin{table}
\centering
\begin{tabular}{cccccc}
\hline\hline
Regime&Method& $\langle Q_2 \rangle_\Gamma$ of $C_p$& $\langle NRMSE
\rangle_\Gamma$ of $C_p$ & $\langle Q_2 \rangle_\Gamma$ of $C_f$&
$\langle NRMSE \rangle_\Gamma$ of $C_f$ \\\hline Both&Classical& 0.897&
1.93$\times 10^{-2}$& 0.837& 1.15$\times 10^{-2}$\\
&LDM&  0.989& 7.48$\times 10^{-3}$& 0.908& 9.53$\times 10^{-3}$\\
Subsonic&Classical& 0.984& 7.46$\times 10^{-3}$& 0.912& 6.61$\times 10^{-3}$\\
&LDM&  0.990& 6.31$\times 10^{-3}$& 0.957& 6.72$\times 10^{-3}$\\
Transonic&Classical& 0.0974& 4.64$\times 10^{-2}$& -0.303& 2.54$\times
10^{-2}$\\
&LDM&  0.901& 9.79$\times 10^{-3}$& -0.0223& 1.72$\times 10^{-2}$\\
\hline\hline
\end{tabular}
\caption{\label{tab:rae_resultsy} Summary of the results in term of $Q_2$ and
NRMSE for the RAE2822.}
\label{table:rae_surrogate_results}
\end{table}
 
\section{Conclusion} \label{conclusion}

In this paper, a local approach for Reduced Order Modeling, called Local
Decomposition Method (LDM), has been presented. It has been developed in order
to cope with problems involving discontinuities and different physical regimes,
common in aerodynamics. The original strategy proposed by the LDM consists in
building a local model for each physical regime identified with machine
learning. Two major steps are associated with this strategy: the use of a
Jameson's shock sensor enhancing the physical regime recognition and an
active sampling adding automatically extra information to the subspace with
the highest nonlinear structures. The LDM has been assessed on a analytical
moving shock problem and the simulation of a turbulent flow around the transonic
RAE2822 airfoil. The results reveal a significant improvement of the model
accuracy, especially in the nonlinear regions.

Further work is needed to increase the efficiency of the strategy if 3D complex
configurations have to be considered. Firstly, the resampling technique has to
be improved. The snapshots added to a specific subspace come from the continuation
of a low-discrepancy sequence. For example, a method minimizing the variance of
the Gaussian Process Regression could be considered. Moreover, the new snapshots
are computed sequentially as the probability of belonging to each subset is
updated at each iteration. A new process of parallelization should be devised to
take advantage of high performance computing during the iterative sampling step
by combining multiple simultaneous jobs into large ensembles. Secondly, the
extrapolation at the interface of the input space parameter remains an open issue.

The extension of this work should be the application of the LDM on
three-dimensional flows at transonic speeds for aerodynamic applications and
aerothermal coupling. It is planned to evaluate this method in a mission
analysis context, taking advantage of the results obtained for the transonic
regions. In particular, the LDM should be used to predict the thermal behavior
of the flow around an engine pylon during different flight phases.

Finally, this work can be seen as a contribution to the coupling of machine
learning and Computational Fluid Dynamics, in particular for the prediction of a
quantity of interest in the case of complex flow fields.


\section*{Acknowledgments} 
This work is part of the MDA-MDO project of the French Institute of Technology
IRT Saint Exupery. We wish to acknowledge the PIA framework (CGI, ANR) and the
project industrial members for their support, financial backing and/or own
knowledge: Airbus, Airbus Group Innovations, SOGETI High Tech, Altran
Technologies, CERFACS.

\section*{References}

\bibliographystyle{aiaa}  
\bibliography{references}

\end{document}